\documentclass[sigconf,9pt,screen]{acmart}
\PassOptionsToPackage{hyphens}{url}\usepackage{hyperref}
\usepackage[english]{babel}
\usepackage{graphicx}
\usepackage{xspace}
\usepackage{booktabs}
\usepackage{xfp}
\usepackage{siunitx}
\usepackage{listings}
\usepackage{algorithmicx}
\usepackage[noend]{algpseudocode}
\usepackage{multirow}
\usepackage[inline]{enumitem}
\usepackage{listings}
\usepackage[normalem]{ulem} %
\usepackage{comment}
\usepackage{caption}
\usepackage{subcaption}
\usepackage{tikz}
\usepackage{gnuplot-lua-tikz}

\microtypecontext{spacing=nonfrench}

\acmYear{2022}\copyrightyear{2022}
\acmConference[SIGCOMM '22]{ACM SIGCOMM 2022 Conference}{August 22--26, 2022}{Amsterdam, Netherlands}
\acmBooktitle{ACM SIGCOMM 2022 Conference (SIGCOMM '22), August 22--26, 2022, Amsterdam, Netherlands}
\acmDOI{10.1145/3544216.3544253}
\acmISBN{978-1-4503-9420-8/22/08}
\setcopyright{cc}
\setcctype{by}



\usepackage{titlesec}
\titlespacing*{\paragraph}{9pt}{1.4mm}{1.4mm}

\def\Snospace~{\S{}}

\usepackage{xcolor}

\newcommand{\sysname}{SimBricks\xspace}
\newcommand{\eg}{e.g.\xspace}

\newcommand{\nd}{\texttt{NetDevice}\xspace}

\renewcommand*\Call[2]{\textsc{#1}(#2)}

\newsavebox{\protobox}

\begin{document}
\fancypagestyle{firstpagestyle}{
  \addtolength{\headheight}{.6in}
  \addtolength{\topmargin}{-.6in}
  \fancyhead[L]{\href{https://simbricks.github.io/}
    {\includegraphics[width=1.2in]{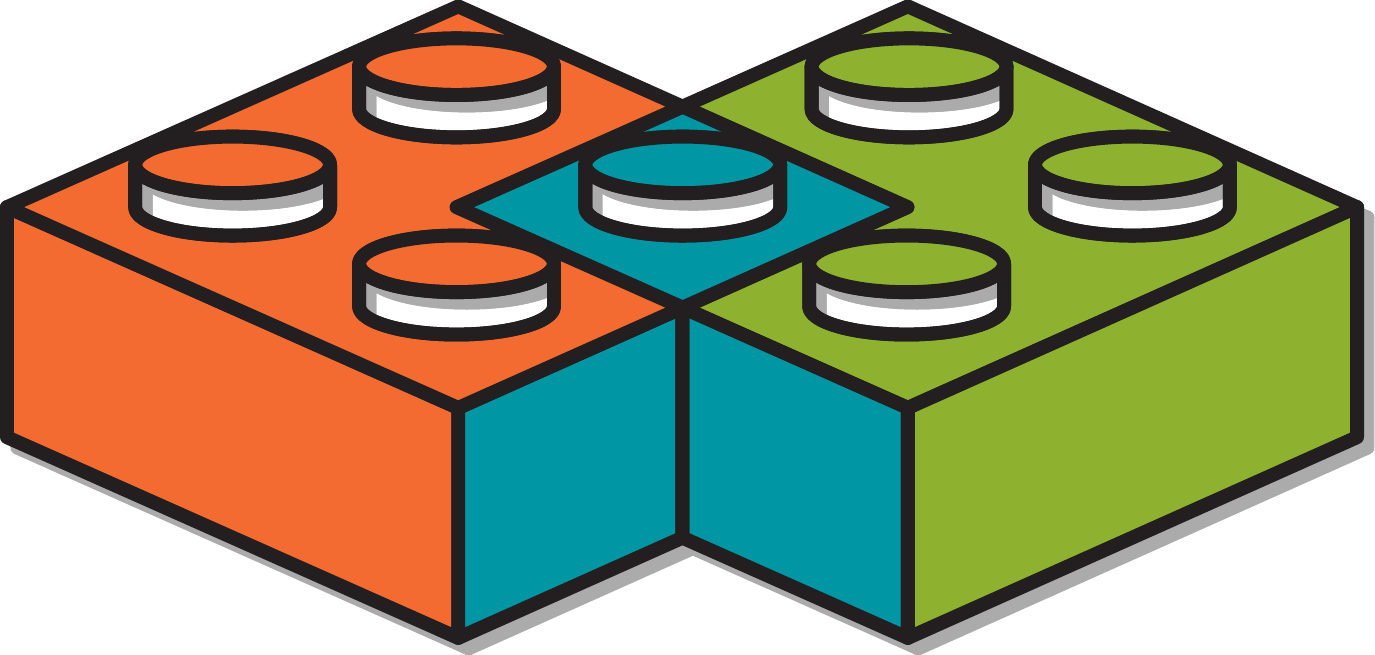}}}
  \fancyhead[R]{\href{https://conferences.sigcomm.org/sigcomm/2022/\#process}
    {\includegraphics[width=2in]{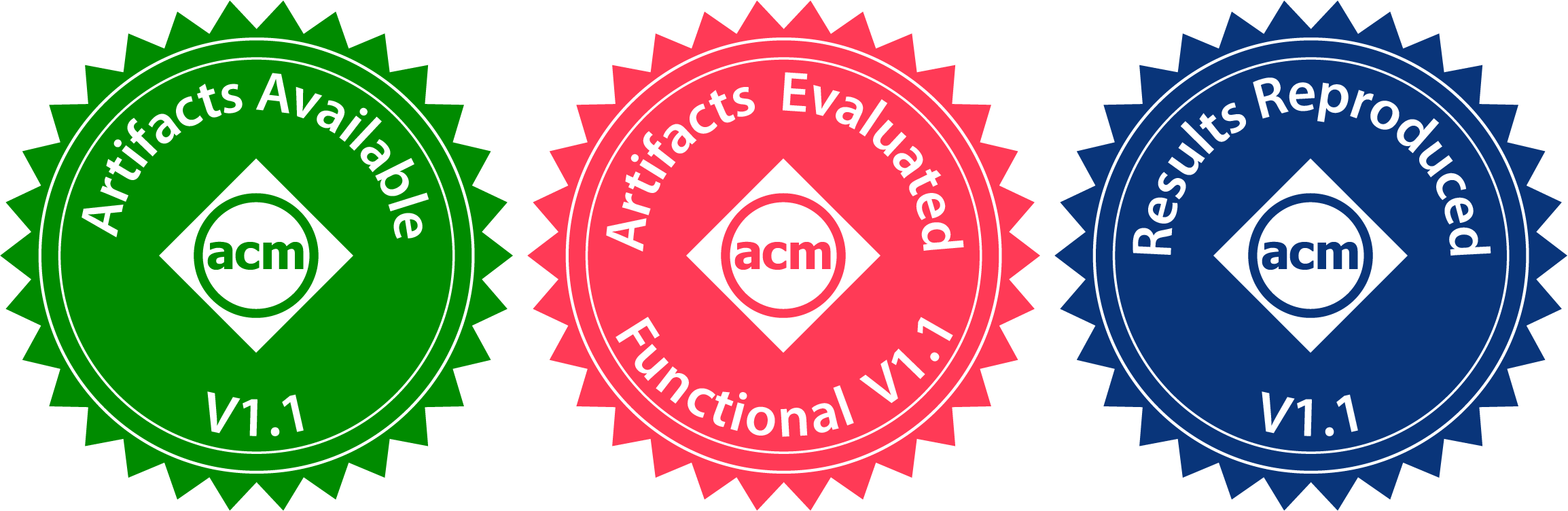}}}
}

\title[\sysname: End-to-End Network System Evaluation with Modular
Simulation]{\sysname: End-to-End Network System Evaluation\\ with Modular
  Simulation}

\author{Hejing Li}
\orcid{0000-0001-6930-2419}
\affiliation{%
\institution{Max Planck Institute for Software Systems (MPI-SWS)}%
\city{Saarbr\"{u}cken}%
\country{Germany}}
\email{hejingli@mpi-sws.org}

\author{Jialin Li}
\orcid{0000-0003-3530-7662}
\affiliation{%
\institution{National University of Singapore}%
\country{Singapore}}
\email{lijl@comp.nus.edu.sg}

\author{Antoine Kaufmann}
\orcid{0000-0002-6355-2772}
\affiliation{%
\institution{Max Planck Institute for Software Systems (MPI-SWS)}%
\city{Saarbr\"{u}cken}%
\country{Germany}}
\email{antoinek@mpi-sws.org}

\begin{abstract}
  Full system ``end-to-end'' measurements in physical testbeds are the
  gold standard for network systems evaluation but are often not
  feasible.
  When physical testbeds are not available we frequently turn to
  simulation for evaluation.
  Unfortunately, existing simulators are insufficient for end-to-end
  evaluation, as they either cannot simulate all components, or
  simulate them with inadequate detail.

  We address this through modular simulation, flexibly
  combining and connecting multiple existing simulators for
  different components, including processor and memory, devices, and
  network, into virtual end-to-end testbeds tuned for each use-case.
  Our architecture, \sysname, combines well-defined component
  interfaces for extensibility and modularity,
  efficient communication channels for local and distributed
  simulation,
  and a co-designed efficient synchronization mechanism for
  accurate timing across simulators.
  We demonstrate \sysname scales to 1000 simulated hosts,
  each running a full software stack including Linux, and that
  it can simulate testbeds with existing NIC and switch RTL
  implementations.
  We also reproduce key findings from prior work in congestion
  control, NIC architecture, and in-network computing in \sysname.
\end{abstract}
 
\maketitle

\section{Introduction}
Our community expects research ideas to be implemented and evaluated as part of
a complete system ``end-to-end'' in realistic conditions.
End-to-end evaluation is important as many factors in each system component
affect the overall behavior in subtle and unpredictable ways.

Yet evaluation in full physical testbeds is frequently infeasible.
Work might require cutting edge commercial hardware that is not yet
available at the time of
publication~\cite{sharma:flexswitch,liu:incbricks,li:nopaxos,eris},
develop hardware extensions to existing proprietary
hardware~\cite{sharma:afq}, or propose entirely new ASIC hardware
architectures~\cite{bosshart:rmt,chole:drmt,sivaraman:packet_transactions,
sutherland:nebula,kaufmann:flexnic,jouppi:tpu,magaki:asic_clouds,
lin:panic}.
The trend towards increasingly specialized hardware, including
SmartNICs, programmable switches, and other accelerators, further
exacerbates this.
Finally, work on network protocols and congestion control necessitates
evaluation in large scale networks with hundreds to thousands of hosts.

When a full evaluation in a physical testbed is not possible,
simulation has long offered an alternative.
In networking, we use ns-2~\cite{software:ns2},
ns-3~\cite{software:ns3}, and OMNeT++~\cite{varga:omnetpp} to evaluate
protocols and algorithms;
computer architects rely on system simulators such as
gem5~\cite{binkert:gem5},
while hardware designers employ RTL simulators such as
Modelsim~\cite{software:modelsim} or
Verilator~\cite{software:verilator}.
While network systems do benefit from these
simulators~\cite{arashloo:tonic,mittal:irn,kaufmann:tas}, they do not
enable end-to-end evaluation, as no existing simulator simulates all
required components in a testbed: hosts, devices, and the full network.

In this paper, we demonstrate how to enable end-to-end network system simulation
by combining different simulators to cover the necessary functionality.
Instead of building a new simulator, throwing away decades of work, we connect
existing and new simulators -- for hosts, hardware devices, and networks -- into
full system simulations capable of running unmodified operating systems,
drivers, and applications.
Existing simulators, however, are standalone and not designed to be
combined with other simulators.
To achieve modular end-to-end simulation, we thus need to overcome
three technical challenges:
1) no interfaces to connect simulators together,
2) efficient, scalable, and correct synchronization of simulator clocks, and
3) combining mutually incompatible simulation models.

We present the design and implementation of \emph{\sysname, a modular
simulation framework for end-to-end network system simulations}.
\sysname defines interfaces for interconnecting simulators based on
natural component boundaries in physical systems, specifically
PCIe and Ethernet links.
Individual component simulators run in \emph{parallel} as separate processes,
and communicate via message passing only between connected peers through
optimized shared memory queues.
With this message transport, we co-design a protocol that leverages
simulation topology and latency at component boundaries for
\emph{efficient and accurate synchronization} of simulator clocks.
For scaling out simulations across physical hosts, we introduce
a proxy to forward messages over TCP or RDMA.

Currently, \sysname integrates QEMU~\cite{software:qemu} and
gem5~\cite{binkert:gem5} as host simulators,
Verilator~\cite{software:verilator} as an RTL hardware simulator for
hardware devices,
and ns-3~\cite{software:ns3}, OMNeT++~\cite{varga:omnetpp}, as well
as the Intel Tofino simulator~\cite{software:p4studio} for
network simulation.
Further, we have integrated open source RTL designs for the Corundum
FPGA NIC~\cite{forencich:corundum} and the
Menshen switch pipeline~\cite{wang:menshen} to showcase \sysname{}'s
generality.
We have also implemented fast behavioral simulators, \eg for
the Intel X710 40G NIC~\cite{spec:intel_x710}, and ported the
FEMU NVMe SSD model~\cite{li:femu} into \sysname.
In combination, these simulators enable a broad range of end-to-end
configurations for different use-cases.

In our evaluation, we demonstrate that \sysname enables end-to-end simulation of
existing network systems at small and large scales.
We also reproduce key results from congestion
control~\cite{alizadeh:dctcp},
in-network compute~\cite{li:nopaxos},
and FPGA NIC design~\cite{forencich:corundum} in \sysname.
\sysname obtains more realistic results compared to ns-3 in isolation
(\S\ref{ssec:bg:sysresearch}).
\sysname also scales to 1000 hosts and NICs with only a
14\% increase in simulation time compared to a 40-host simulation
(\S\ref{ssec:eval:scalability}).
Finally, \sysname provides deep visibility and control of low-level system
behaviors, facilitating evaluation and performance debugging
(\S\ref{ssec:eval:corundum}).

\smallskip\noindent
We make the following technical contributions:
\begin{itemize}
  \item \emph{Modular architecture for end-to-end system simulation}
    (\S\ref{ssec:design:interface}) combining host, device, and network
    simulators.

  \item \emph{Co-designed message transport and synchronization
    mechanism for parallel and distributed simulations}
    (\S\ref{ssec:design:syncproto}, \S\ref{ssec:design:transport}) leveraging
    pairwise message passing to efficiently ensure correct simulation, even
    at scale.

  \item \sysname, a \emph{prototype implementation} of our
    architecture (\S\ref{sec:impl}) with integrations for
    existing and new simulators.
\end{itemize}

\noindent
\sysname is available open source at \url{https://simbricks.github.io}

\noindent
This work does not raise any ethical issues.

\section{Simulation Background}
Simulators employ techniques such as discrete event
simulation, binary translation, and hardware virtualization, to
simulate system components at various scales and levels of detail.
Network simulators, such as ns-2~\cite{software:ns2},
ns-3~\cite{software:ns3}, and OMNeT++~\cite{varga:omnetpp}, use
discrete event simulation to model packets traversing network
topologies.
Computer architecture simulators, such as gem5~\cite{binkert:gem5},
QEMU~\cite{software:qemu}, and Simics~\cite{magnusson:simics},
simulate full computer systems capable of running unmodified guest
software, including operating systems, with different and sometimes
configurable degrees of detail.
These simulators also include I/O devices, but often only implement
the minimum features for basic functionality.
Hardware RTL simulations, such as xsim~\cite{software:vivado_sim} and
Verilator~\cite{software:verilator}, help test and debug hardware
designs cycle by cycle against testbenches.
In all three cases
\emph{individual components are simulated in isolation}.

\paragraph{Advantages.}
The main motivation for simulation is that a physical implementation is often
not feasible.
Simulations are also \emph{portable}
as they decouple the simulated system from the host system.
Many are deterministic (with explicit seeds for randomness), providing
\emph{reproducible results}.
Simulators are also \emph{flexible}; implemented as software they can
be modified, and frequently offer parameters representing a broad
range of configurations.
Finally, simulations provide great \emph{visibility}, and can log
details about the system, without affecting behavior.

\paragraph{Disadvantages.}
Simulations also have some common drawbacks.
\emph{Long simulation times} are common -- architectural simulators
often only simulate hundreds or thousands of system cycles a
second~\cite{sutherland:nebula,karandikar:firesim}, and simulating a
few milliseconds of a large scale topology in ns-3 can take many
hours.
Different simulators strike different trade-offs between accuracy and
simulation time, depending on the intended use-case.
Finally, simulation results are only as good as the simulator, and may
not be representative unless \emph{validated} against a physical
testbed.

\paragraph{Comparison to Emulation.}
Emulations replicate externally visible behavior of a system without
modeling internal details, and typically run at close to interactive
speeds.
For example, Mininet~\cite{lantz:mininet} emulates OpenFlow networks with
multiple end-hosts running real Linux applications at near native speed on a
single physical host, by using Linux containers and kernel network features.
However, as emulation uses wall-clock time, it only works as long as
all components can keep up in real time.
Simulations, in contrast, rely on virtual time which can slow down without
affecting simulated behavior.
Additionally, emulation does not model internals of a system that could affect
system behavior, e.g., interactions between NIC and drivers.
As such, emulation is primarily useful for interactive testing or performance
evaluation when fidelity is not crucial.

\section{Systems Research Challenges}
\label{ssec:bg:sysresearch}
Systems research faces additional challenges that complicate using
simulation during prototyping and evaluation.

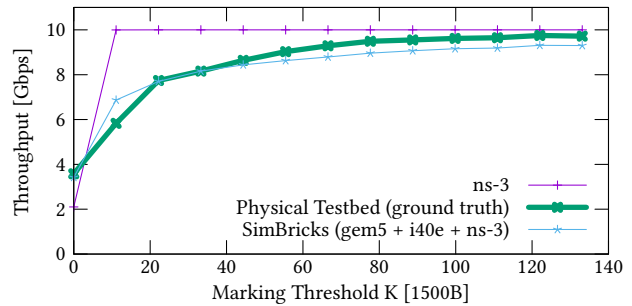
\begin{figure}%
\centering%
\begin{tikzpicture}[gnuplot]
\tikzset{every node/.append style={font={\fontsize{8.0pt}{9.6pt}\selectfont}}}
\path (0.000,0.000) rectangle (8.458,4.216);
\gpcolor{color=gp lt color border}
\gpsetlinetype{gp lt border}
\gpsetdashtype{gp dt solid}
\gpsetlinewidth{1.00}
\draw[gp path] (0.907,0.688)--(1.087,0.688);
\draw[gp path] (8.016,0.688)--(7.836,0.688);
\node[gp node right] at (0.907,0.688) {$0$};
\draw[gp path] (0.907,1.285)--(1.087,1.285);
\draw[gp path] (8.016,1.285)--(7.836,1.285);
\node[gp node right] at (0.907,1.285) {$2$};
\draw[gp path] (0.907,1.881)--(1.087,1.881);
\draw[gp path] (8.016,1.881)--(7.836,1.881);
\node[gp node right] at (0.907,1.881) {$4$};
\draw[gp path] (0.907,2.478)--(1.087,2.478);
\draw[gp path] (8.016,2.478)--(7.836,2.478);
\node[gp node right] at (0.907,2.478) {$6$};
\draw[gp path] (0.907,3.074)--(1.087,3.074);
\draw[gp path] (8.016,3.074)--(7.836,3.074);
\node[gp node right] at (0.907,3.074) {$8$};
\draw[gp path] (0.907,3.671)--(1.087,3.671);
\draw[gp path] (8.016,3.671)--(7.836,3.671);
\node[gp node right] at (0.907,3.671) {$10$};
\draw[gp path] (0.907,0.688)--(0.907,0.868);
\draw[gp path] (0.907,3.969)--(0.907,3.789);
\node[gp node center] at (0.907,0.515) {$0$};
\draw[gp path] (1.923,0.688)--(1.923,0.868);
\draw[gp path] (1.923,3.969)--(1.923,3.789);
\node[gp node center] at (1.923,0.515) {$20$};
\draw[gp path] (2.938,0.688)--(2.938,0.868);
\draw[gp path] (2.938,3.969)--(2.938,3.789);
\node[gp node center] at (2.938,0.515) {$40$};
\draw[gp path] (3.954,0.688)--(3.954,0.868);
\draw[gp path] (3.954,3.969)--(3.954,3.789);
\node[gp node center] at (3.954,0.515) {$60$};
\draw[gp path] (4.969,0.688)--(4.969,0.868);
\draw[gp path] (4.969,3.969)--(4.969,3.789);
\node[gp node center] at (4.969,0.515) {$80$};
\draw[gp path] (5.985,0.688)--(5.985,0.868);
\draw[gp path] (5.985,3.969)--(5.985,3.789);
\node[gp node center] at (5.985,0.515) {$100$};
\draw[gp path] (7.000,0.688)--(7.000,0.868);
\draw[gp path] (7.000,3.969)--(7.000,3.789);
\node[gp node center] at (7.000,0.515) {$120$};
\draw[gp path] (8.016,0.688)--(8.016,0.868);
\draw[gp path] (8.016,3.969)--(8.016,3.789);
\node[gp node center] at (8.016,0.515) {$140$};
\draw[gp path] (0.907,3.969)--(0.907,0.688)--(8.016,0.688)--(8.016,3.969)--cycle;
\node[gp node center,rotate=-270] at (0.232,2.328) {Throughput [Gbps]};
\node[gp node center] at (4.461,0.171) {Marking Threshold K [1500B]};
\node[gp node right] at (6.807,1.605) {ns-3};
\gpcolor{rgb color={0.580,0.000,0.827}}
\draw[gp path] (6.954,1.605)--(7.722,1.605);
\draw[gp path] (0.907,1.315)--(1.470,3.670)--(2.034,3.671)--(2.597,3.671)--(3.160,3.671)%
  --(3.724,3.671)--(4.287,3.671)--(4.850,3.671)--(5.413,3.671)--(5.977,3.671)--(6.540,3.671)%
  --(7.103,3.671)--(7.667,3.671);
\gpsetpointsize{4.00}
\gp3point{gp mark 1}{}{(0.907,1.315)}
\gp3point{gp mark 1}{}{(1.470,3.670)}
\gp3point{gp mark 1}{}{(2.034,3.671)}
\gp3point{gp mark 1}{}{(2.597,3.671)}
\gp3point{gp mark 1}{}{(3.160,3.671)}
\gp3point{gp mark 1}{}{(3.724,3.671)}
\gp3point{gp mark 1}{}{(4.287,3.671)}
\gp3point{gp mark 1}{}{(4.850,3.671)}
\gp3point{gp mark 1}{}{(5.413,3.671)}
\gp3point{gp mark 1}{}{(5.977,3.671)}
\gp3point{gp mark 1}{}{(6.540,3.671)}
\gp3point{gp mark 1}{}{(7.103,3.671)}
\gp3point{gp mark 1}{}{(7.667,3.671)}
\gp3point{gp mark 1}{}{(7.338,1.605)}
\gpcolor{color=gp lt color border}
\node[gp node right] at (6.807,1.310) {Physical Testbed (ground truth)};
\gpcolor{rgb color={0.000,0.620,0.451}}
\gpsetlinewidth{5.00}
\draw[gp path] (6.954,1.310)--(7.722,1.310);
\draw[gp path] (0.907,1.756)--(1.470,2.427)--(2.034,2.997)--(2.597,3.122)--(3.160,3.265)%
  --(3.724,3.381)--(4.287,3.459)--(4.850,3.519)--(5.413,3.537)--(5.977,3.557)--(6.540,3.566)%
  --(7.103,3.596)--(7.667,3.587);
\gp3point{gp mark 2}{}{(0.907,1.756)}
\gp3point{gp mark 2}{}{(1.470,2.427)}
\gp3point{gp mark 2}{}{(2.034,2.997)}
\gp3point{gp mark 2}{}{(2.597,3.122)}
\gp3point{gp mark 2}{}{(3.160,3.265)}
\gp3point{gp mark 2}{}{(3.724,3.381)}
\gp3point{gp mark 2}{}{(4.287,3.459)}
\gp3point{gp mark 2}{}{(4.850,3.519)}
\gp3point{gp mark 2}{}{(5.413,3.537)}
\gp3point{gp mark 2}{}{(5.977,3.557)}
\gp3point{gp mark 2}{}{(6.540,3.566)}
\gp3point{gp mark 2}{}{(7.103,3.596)}
\gp3point{gp mark 2}{}{(7.667,3.587)}
\gp3point{gp mark 2}{}{(7.338,1.310)}
\gpcolor{color=gp lt color border}
\node[gp node right] at (6.807,1.015) {\sysname (gem5 + i40e + ns-3)};
\gpcolor{rgb color={0.337,0.706,0.914}}
\gpsetlinewidth{1.00}
\draw[gp path] (6.954,1.015)--(7.722,1.015);
\draw[gp path] (0.907,1.708)--(1.470,2.737)--(2.034,2.985)--(2.597,3.122)--(3.160,3.205)%
  --(3.724,3.262)--(4.287,3.310)--(4.850,3.361)--(5.413,3.393)--(5.977,3.420)--(6.540,3.429)%
  --(7.103,3.465)--(7.667,3.462);
\gp3point{gp mark 3}{}{(0.907,1.708)}
\gp3point{gp mark 3}{}{(1.470,2.737)}
\gp3point{gp mark 3}{}{(2.034,2.985)}
\gp3point{gp mark 3}{}{(2.597,3.122)}
\gp3point{gp mark 3}{}{(3.160,3.205)}
\gp3point{gp mark 3}{}{(3.724,3.262)}
\gp3point{gp mark 3}{}{(4.287,3.310)}
\gp3point{gp mark 3}{}{(4.850,3.361)}
\gp3point{gp mark 3}{}{(5.413,3.393)}
\gp3point{gp mark 3}{}{(5.977,3.420)}
\gp3point{gp mark 3}{}{(6.540,3.429)}
\gp3point{gp mark 3}{}{(7.103,3.465)}
\gp3point{gp mark 3}{}{(7.667,3.462)}
\gp3point{gp mark 3}{}{(7.338,1.015)}
\gpcolor{color=gp lt color border}
\draw[gp path] (0.907,3.969)--(0.907,0.688)--(8.016,0.688)--(8.016,3.969)--cycle;
\gpdefrectangularnode{gp plot 1}{\pgfpoint{0.907cm}{0.688cm}}{\pgfpoint{8.016cm}{3.969cm}}
\end{tikzpicture}
\caption{Throughput for two dctcp flows in ns-3, a physical
      testbed, and a \sysname end-to-end simulation.}%
\label{fig:dctcp}%
\Description{Line graph showing the throughput on the y-axis and the
  dctcp marking threshold K for three configurations: a physical
  testbed (the ground thruth), a conventional network-only ns-3
  simulation, and \sysname (with gem5, i40e, and ns-3). The lines for
  the physical testbed and \sysnets match closely with just minor
  discrepancies. ns-3 on the other hand immediately reaches line-rate
  at K=10, where the physical testbed still only manages 6\,Gbps.}
\end{figure}

\paragraph{Not end-to-end.}
First and foremost, \emph{no existing simulator covers all
required components for network systems with sufficient features and
detail}, precluding end-to-end evaluation.
While existing simulators cover individual components, such as computer
architecture, hardware devices, and networks, they only do so in isolation with
no mechanism for combining them into complete systems.
As a result, we are left with non-end-to-end ``piecemeal'' evaluation,
where different components are evaluated in isolation~\cite{
  arashloo:tonic,mittal:irn,handley:ndp}.

We illustrate the pitfalls of piecemeal evaluation by comparing
dctcp~\cite{alizadeh:dctcp} congestion control behavior
in the ns-3 network simulator to a physical testbed.
As network speed increases and bottlenecks move to end-hosts,
congestion control incurs small variations in timing in the host
hardware and software which can affect
behavior~\cite{alizadeh:dctcp,mittal:timely,kumar:swift}.
However, ns-3 only models network and protocol behavior, and as
a result, does not capture these factors.
We set up two clients and two servers sharing a single 10G
bottleneck link with a 4000B MTU, and one large TCP flow
generated by iperf for each client-server pair.
\autoref{fig:dctcp} shows the throughput for varying dctcp
marking thresholds $K$.
The marking threshold balances queuing latency and throughput; a lower
threshold reduces queue length but risks under-utilizing links.
ns-3 underestimates the necessary threshold~\cite{alizadeh:dctcp} to
achieve line rate, as it does not model host processing variations,
particularly processing delay caused by OS interrupt scheduling.
Only an end-to-end evaluation of the full system captures such
intricacies.

\paragraph{Not scalable.}
Network and distributed systems frequently require evaluation on clusters beyond
tens of hosts to demonstrate scalability.
But for most simulators, already long simulation times increase super-linearly
with the size of the simulated system, making simulation of a large network
system an infeasible task.

\paragraph{Not modular.}
Using simulators for systems research often requires extending
existing simulators with additional functionality, \eg, adding a new
NIC to an architecture simulator.
These extensions are tied to a particular simulator, as different
simulators lack common internal interfaces,
This complicates apples-to-apples comparisons for future work that may use a
different simulator, \eg, to simulate a host with a different NIC, forcing the
same simulator to be used throughout the project cycle.
Finally, this tight integration complicates the implementation and releasing of
such extensions, as they often require maintaining a fork of the full simulator.
\section{Modular Simulation}
\label{sec:approach}
We argue that \emph{end-to-end simulations can be effectively
assembled from multiple different interconnected and synchronized
simulators for individual components.}
To demonstrate this, we present \sysname, a new modular simulation
framework that aims to provide end-to-end network system simulation.

\begin{figure}[t]%
\centering%
\includegraphics[width=\columnwidth]{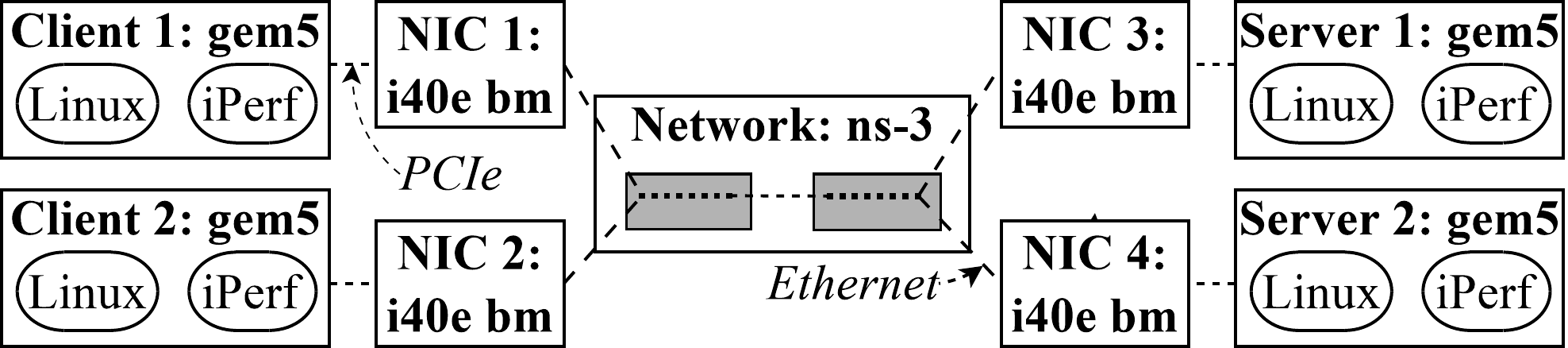}%
\caption{\sysname configuration for the dctcp experiment in
  \autoref{fig:dctcp}, combining gem5, ns-3, and an Intel NIC
  simulator. Each simulator runs in a separate process.}%
\label{fig:dctcp-setup}%
\Description{Digram with boxes for individual simulators and their
  connections. Starting on the left, two boxes representing separate
  gem-5 instances running Linux and iPerf as the client hosts. These
  connect to a separate next box for an i40e NIC simulator instance
  each. Finally these client NICs connect to a central box that
  represents the ns-3 network simulator. On the right side the
  mirror-image for two server hosts, again with two NICs and two
  servers.}
\end{figure}

\paragraph{End-to-end simulations are better.}
Returning to the dctcp example from earlier, \autoref{fig:dctcp-setup}
shows the simulation setup that produces the result shown in
\autoref{fig:dctcp}.
We combine four instances of gem5 with four instances of the Intel
i40e NIC simulator we developed, each pair connected through PCIe;
all NIC simulators are in turn connected to an instance of ns-3.
The gem5 instances are running a full Ubuntu image with
unmodified NIC drivers and iperf.
\autoref{fig:dctcp} shows that our \sysname simulation approximates
the behavior of the physical testbed much more closely than ns-3, and
yields the same insight.
We conclude that end-to-end evaluation with \sysname improves accuracy
for network system evaluation over non-end-to-end simulators.

\subsection{Design Goals}
\label{ssec:approach:goals}
To address the challenges for using simulations in systems research,
(\autoref{ssec:bg:sysresearch}), we have the following design goals
for
\sysname:

\begin{itemize}
  \item \textbf{End-to-end}: simulate full network systems,
    with hosts, existing or custom devices, network topologies, and
    the full software stack, including unmodified OS and applications.
  \item \textbf{Scalable}: simulate large network systems consisting
    of tens or hundreds of separate hosts and devices.
  \item \textbf{Fast}: keep simulation times as low as possible.
  \item \textbf{Modular}: enable flexible composition of simulators, where
    components can be added and swapped independently.
  \item \textbf{Accurate}: preserve accuracy of constituent
    simulators, correctly interface and synchronize components to
    behave equivalent to a monolithic simulator with the same
    models.
  \item \textbf{Deterministic}: keep end-to-end simulation
    deterministic when all individual simulators are deterministic.
  \item \textbf{Transparent}: provide deep and detailed visibility
    into end-to-end performance without affecting simulation behavior,
    to support debugging and performance analysis.
\end{itemize}

\subsection{Technical Challenges}
Achieving our design goals incurs the following challenges:

\paragraph{Simulation interconnection interfaces.}
Unfortunately, existing simulators are standalone and provide no suitable
interfaces for interconnecting with other external simulators.
Moreover, enabling modular ``plug-and-play'' configurations, where
components can be independently swapped out, requires common,
well-defined interfaces between different component types.

\paragraph{Scalable synchronization and communication.}
Individual component simulators maintain their own virtual simulation
clocks that progress at different rates.
To accurately connect simulators, we need to synchronize their virtual
clocks.
However, this synchronization comes at a performance cost, especially
with increasing system scale.
For example, we measure a $3.7\times$ increase in runtime for the
dist-gem5~\cite{mohammad:distgem5} simulator when scaling from 2 to 16
simulated hosts, due to synchronization overhead
(\autoref{ssec:eval:syncproto}).
Prior work shows synchronization overhead can be reduced by
sacrificing accuracy and determinism through lax synchronization.
\cite{fujimoto:relaxedsim,chen:slacksim}.
Since this violates two of our design goals, we do not consider this.

\paragraph{Incompatible simulation models.}
Finally, different simulators often employ mutually incompatible
simulation models.
For example, QEMU has a synchronous device model where calls in device
code block until complete, while ns-3 schedules asynchronous events
to model networks, and Verilator simulates hardware circuits cycle by
cycle.
We therefore need an interface compatible with all of these simulation
models.

\subsection{Design Principles}
We address these challenges through four design principles:

\paragraph{Fix natural component simulator interfaces.}
To enable modular composition of simulators, \sysname defines an
interface for each \textit{component type}
(\autoref{ssec:design:interface}).
We base these interfaces on the \emph{point-to-point} component
boundaries in real systems:
PCI express (PCIe) connects today's hardware devices to servers, while
network devices typically connect through Ethernet networks.
We choose these interfaces as a starting point, but our approach
generalizes to other interconnects and networks.
These component interfaces form narrow waists, decoupling innovation on
both sides:
To integrate a simulator into \sysname, developers need to add an
adapter that implements the component interface, without needing to
modify other simulators.
We assume a static topology of components throughout a simulation.

\paragraph{Loose coupling with message passing.}
Instead of tightly integrating multiple simulators into one simulation
loop, \sysname runs component simulators as separate processes that
communicate through message passing (\autoref{ssec:design:interface})
across our defined interfaces.
This drastically simplifies integrating simulators into \sysname,
as we treat each simulator as a black-box that only needs to
implement our interfaces.
Using asynchronous message passing also maximizes compatibility with
different simulation models:
Discrete event and cycle-by-cycle simulations can issue requests and
process responses at the scheduled times, while blocking simulations
can block till the response message arrives --- for peer simulators
this is transparent.
Message passing channels also provide inspection points for debugging
and tracing system behavior without modifying component simulators.

\paragraph{Parallel execution with shared memory queues.}
We run simulators in parallel on different host cores and connect them
through optimized shared-memory queues
(\autoref{ssec:design:transport}).
As simulators run on separate cores and only communicate when
necessary, this avoids unnecessary cache-coherence traffic and
hidden scalability bottlenecks.
These mechanisms allow us to
(i) \textit{scale up} to large simulations: Instead of simulating the
complete system in one simulation instance, we simulate different
components of the system in separate simulators running in parallel
(\autoref{ssec:design:decomp}).
(ii) \textit{scale out} with distributed simulations: We use a
separate proxy that transparently forwards messages on shared memory
queues over the network to and from simulators running on remote hosts
(\autoref{ssec:design:proxy}).

\paragraph{Accurate and efficient synchronization.}
We ensure accurate simulation through correct time synchronization among
simulators, but with minimum runtime overhead.
\emph{Synchronization is optional}, and the user can disable it for
unsynchronized emulations.
For this, we combine three key insights:
1)~\emph{Global synchronization is not necessary} as our simulator
boundaries at point-to-point interfaces limit which simulators
directly communicate.
As long as events at these pairwise interfaces are processed in a
time-synchronized manner, simulation behavior is correct.
2)~\emph{Latency at component interfaces provides slack}, reducing
frequency of component having to wait for others to
coordinate~\cite{chen:slacksim} and thus synchronization overhead.
An event sent at time $T$ only arrives at $T+\Delta$, as our component
interfaces have an inherent latency $\Delta$ in physical systems that
we model.
3)~By \emph{inlining synchronization with efficient polled
message transfers}, synchronization overheads can be minimized and
sometimes completely avoided.
We combine these observations to design an accurate, efficient, and
scalable synchronization mechanism for parallel end-to-end simulations
(\autoref{ssec:design:syncproto}).

\subsection{Non-Goals}
\sysname is not a panacea. We explicitly view the following aspects as
out of scope for this paper and leave them for future work:

\paragraph{Accelerating component simulators.} \sysname does not
generally aim to reduce simulation times for individual component
simulators as we only modify simulators to add \sysname adapters.
Simulation times for synchronized end-to-end \sysname simulations are
at least as high as the slowest component simulator, and may increase
due to synchronization and communication overhead.
However, in a few cases, \sysname interfaces enable developers to
decompose an existing component simulator into multiple smaller
parallel pieces, thereby reducing simulation time
(\autoref{ssec:eval:decomp}).

\paragraph{Avoiding need for validation.}
To obtain representative results, users need to validate component
simulation configurations in \sysname as with any other simulation.
Validation effort is no higher in \sysname than it would be in an
equivalent monolithic simulator, as \sysname forwards timestamped
events accurately from one simulator-internal interface to another
without modifying them (except for the configured link latency).
We expect, however, that \sysname could reduce validation effort by
allowing users to re-combine validated component simulator
configurations without validating from scratch.
(\autoref{sec:discussion})

\paragraph{Interfacing semantically incompatible simulators.}
While \sysname can combine simulators that use different models for
simulation, it cannot bridge semantic gaps between simulators.
For example, \sysname cannot connect a gem5 host sending packets
through an RTL NIC with a flow-based network simulator.
Such conversions may be possible in special cases, but are specific to
the concrete simulators, and as such could be integrated as part of a
\sysname adapter in such a simulator.
\section{Design}
\label{sec:design}

\begin{figure}%
\centering%
\includegraphics[width=0.4\textwidth]{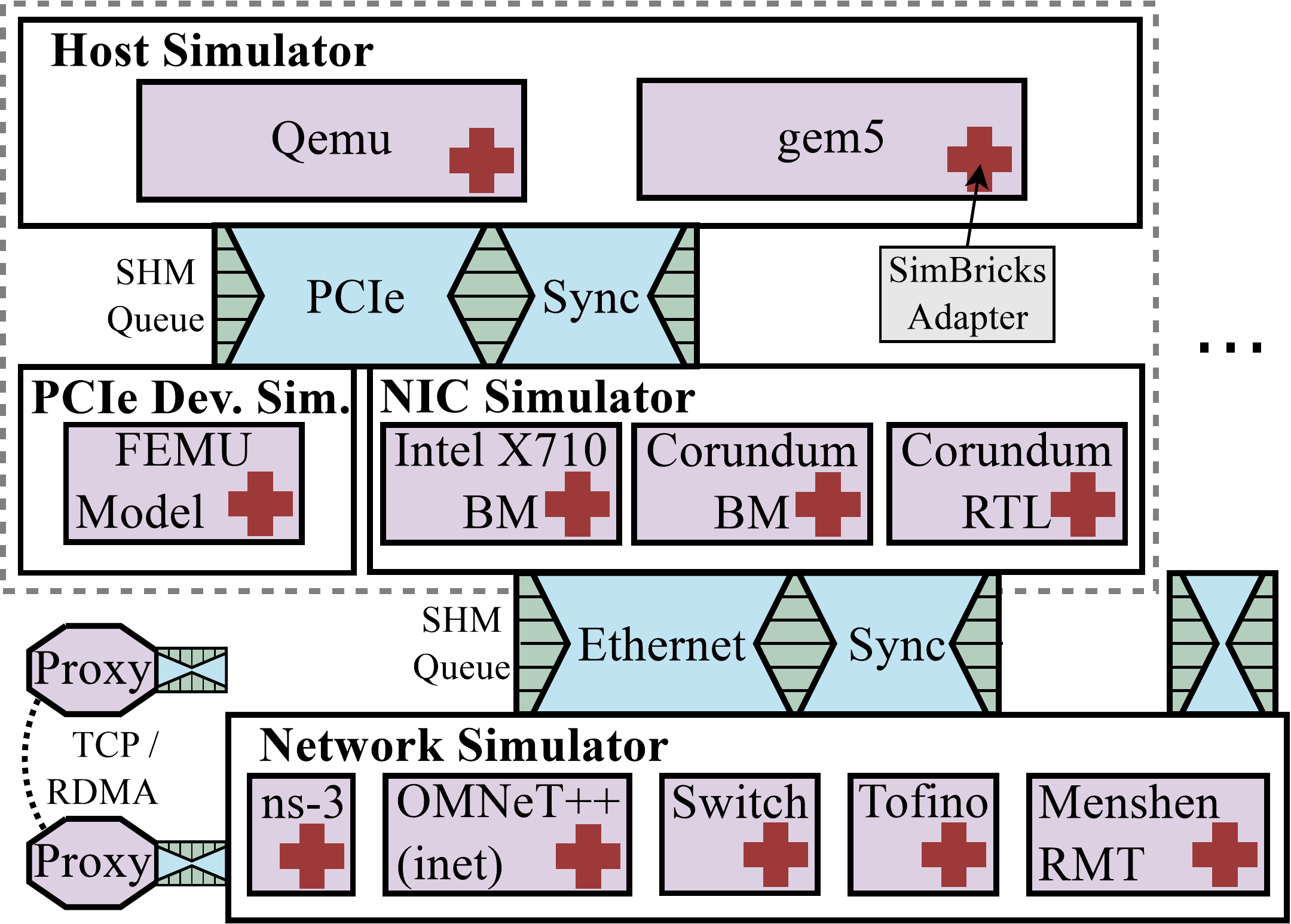}%
\caption{\sysname architecture.
\emph{Double hour glass} with narrow waists between hosts and
  devices, and NICs and networks.}%
\label{fig:sim-arch}%
\Description{Layered architecture diagram. At the two two layers
  grouped loosely: host simulators and PCIe device simulators.
  At the bottom the network simulator layers. host layers connect
  to PCIe device, some of the PCIe deices (NICs) connect to network
  simulators. These connections represented as narrow waists, are
  realized through shared memory queues. RDMA and TCP proxy pairs
  connect to these same connectors. Inside each layer the specific
  simulators along with a symbol indicating a \sysnet adapters are
  shown as boxes. Host simulators are: QEMU and gem5. PCIe devices
  are: the FEMU ssd model, and three NICs (Intel x710 behavioral
  model, Corundum behavioral model, and Corundum RTL simulation).
  Finally network simulators are: ns-3, OMNET++ (inet), switch,
  Tofino, and Menshen RMT.}%
\end{figure}

Using our design principles, we have built \sysname, a modular,
end-to-end simulation framework shown in \autoref{fig:sim-arch}.
In this section, we detail the design of \sysname{}, including
simulator interfaces, fast message transport, techniques to scale
up and out to larger simulations, and the synchronization mechanism.

\subsection{Component Simulator Interfaces}
\label{ssec:design:interface}
\sysname{} achieves modularity through well-defined interfaces
between component simulators:
Host simulators connect to device simulators through a PCIe interface;
NIC and network simulators interconnect through an
Ethernet interface.
This results in a double hourglass architecture
(\autoref{fig:sim-arch}) with narrow waists at component boundaries.
In physical systems both interfaces are asynchronous and incur
propagation delay ($\Delta_i$).
We replicate both aspects.

\begin{figure}[t]%
\centering%
\begin{minipage}{\linewidth}%
\centering%
\small%
\begin{tabular}{cl}
  \toprule[1.5pt]
    \multicolumn{2}{c}{\textbf{PCIe:} Device $\rightarrow$ Host} \\
    \textbf{Message Type} &
    \textbf{Message Fields} \\
  \midrule[1pt]
    \texttt{INIT\_DEV} &
    \begin{tabular}[c]{@{}l@{}}
      PCI vendor, device id, class, \\
      subclass, revision, \\
      \# of MSI vectors, \# of MSI-X vectors, \\
      table/PBA bar and offset
    \end{tabular} \\
  \midrule[0.5pt]
    \begin{tabular}[c]{@{}c@{}}
      \texttt{DMA\_READ},\\
      \texttt{DMA\_WRITE}\\
    \end{tabular} &
    \begin{tabular}[c]{@{}l@{}}
      request ID, memory address, length, \\
      payload data (optional)
    \end{tabular} \\
  \midrule[0.5pt]
    \texttt{MMIO\_COMPL} &
    request ID, payload data (optional) \\
  \midrule[0.5pt]
    \texttt{INTERRUPT} &
    \begin{tabular}[c]{@{}l@{}}
      interrupt type, \\
      MSI/MSI-X: vector \#, \\
      legacy: level
    \end{tabular} \\
  \midrule[0.5pt]
  \midrule[1.5pt]
    \multicolumn{2}{c}{\textbf{PCIe:} Host $\rightarrow$ Device} \\
    \textbf{Message Type} &
    \textbf{Message Fields} \\
  \midrule[1pt]
    \texttt{DMA\_COMPL} &
    request ID, payload data (optional) \\
  \midrule[0.5pt]
    \begin{tabular}[c]{@{}c@{}}
      \texttt{MMIO\_READ}, \\
      \texttt{MMIO\_WRITE}
    \end{tabular} &
    \begin{tabular}[c]{@{}l@{}}
      request ID, BAR \# and offset, length, \\
      payload data (optional)
    \end{tabular} \\
  \midrule[0.5pt]
    \texttt{INT\_STATUS} &
    interrupts enabled: legacy, MSI, MSI-X \\
  \midrule[1.5pt] \\
  \midrule[1.5pt]
    \multicolumn{2}{c}{\textbf{Ethernet:} NIC $\leftrightarrow$
    Net / Net $\leftrightarrow$ Net} \\
    \textbf{Message Type} &
    \textbf{Message Fields} \\
  \midrule[1pt]
    \texttt{PACKET} &
    \begin{tabular}[c]{@{}l@{}}
      packet length, packet data
    \end{tabular} \\
  \bottomrule[1.5pt]
\end{tabular}%
\end{minipage}%
\caption{\textls[-10]{\sysname's simulator interfaces: PCIe between host and
      device, and Ethernet between network components.}}%
\label{fig:interface}%
\Description{Messages for the PCIe and Ethernet protocols in
\sysname. For PCIe the protocol is asymmetrical, while Ethernet is
symmetrical. For PCIe from the device to the host, the messages are:
INIT\_DEV (with PCI vendor, device, class, subclass, revision IDs, the
number of MSI and MSI-X vectors, and the MSI-X table and PBA bars and
offsets), DMA\_READ and DMA\_WRITE (with request ID, memory address,
length, and payload for write), MMIO\_COMPL (with request ID, payload
data for read), and INTERRUPT (interrupt type, MSI/MSI-X vector, or
legacy interrupt level). In the opposite direction, Host to device the
messages are: DMA\_COMPL (with request ID, payload data for reads),
MMIO\_READ and MMIO\_WRITE (with request ID, BAR number and offset,
length, payload data for writes), INT\_STATUS (with flags for enabled
interrupts: legacy, MSI, MSI-X). For Ethernet there is only one
message: PACKET (with length and packet data).}%
\end{figure}

\subsubsection{PCIe: Host-Device Interface}\hfill\\
PCIe itself is a layered protocol, ranging from the
low-level physical layer to the transactional layer for data
operations.
We define \sysname{}'s host-device interface (\autoref{fig:interface})
based on the PCIe \textit{transactional layer}, and abstract
away physical attributes of the PCIe link with simple parameters --
link bandwidth and latency.
Low-level complexity such as encoding and signaling are unnecessary
for most system simulations and would incur substantial cost and
complexity for each simulator.
Should future use-cases need to model this, a detailed PCIe simulator
could be integrated as an interposed component
(\autoref{ssec:design:decomp}).

\paragraph{Discovery and Initialization.}
A key PCIe feature is that hosts can enumerate and identify connected
devices and the features they support.
To this end, our interface defines the \texttt{INIT\_DEV} message for
registering device simulators with the host simulator.
The device simulator includes device information in the message, such as the PCI
vendor, device identifiers, base address registers (BARs), the number of MSI(-X)
interrupt vectors, and addresses of the MSI-X table and PBA.
The host simulator uses this information to expose a corresponding PCIe device
to the system.

\paragraph{Data transfers: MMIO \& DMA.}
PCIe data transfers are symmetrical: both sides can initiate reads
and writes, which the other side completes.
\sysname{}'s PCIe interface defines \texttt{DMA\_READ} /
\texttt{WRITE} messages for DMA transfers initiated by device
simulators, and \texttt{MMIO\_READ} / \texttt{WRITE} for MMIO accesses
initiated by host simulators.
As in PCIe, all data transfer operations are \textit{asynchronous}.
Once a request is finished, the device simulator issues a \texttt{MMIO\_COMPL}
completion message, while the host simulator adapter sends a
\texttt{DMA\_COMPL}.
PCIe allows multiple outstanding operations and only guarantees that
they will be issued to the memory system in the order of arrival.
Completion events, however, may arrive out-of-order.
To match completions with outstanding requests, all requests carry an
identifier that the receiving simulator includes in the response.

\paragraph{Interrupts.}
Our interface supports all PCIe interrupt signaling methods:
legacy interrupts (INTX), message signaled interrupts (MSI), and
MSI-X.
Physical PCIe devices implement MSI (including configuration,
masking, and generating signalling operations) completely on the
device side.
To reduce repeated implementation effort in device simulators and
integration challenges in host simulators, we instead opt to keep this
functionality inside the host simulator.
Device issues \texttt{INTERRUPT} messages to either trigger an
interrupt vector for MSI(-X) or to set interrupt pin state for INTX.
To support devices that require knowledge about which interrupt
mechanisms the OS has enabled,
our interface provides the \texttt{INT\_STATUS} message which the host simulator
sends on configuration changes.

\subsubsection{Ethernet: Network Component Interface} \hfill\\
\label{sssec:design:interface:nicnet}
In \sysname{}'s network interface, we similarly abstract away
low-level details of the Ethernet standard, and only expose Ethernet
frames, as \texttt{PACKET} messages, to NIC and network
simulators.
A \texttt{PACKET} message carries the length of the packet alongside
packet payload, but omits CRCs to reduce overhead as none of our
network simulators models them and most NICs strip them after
validation.
If future network or NIC simulators require CRCs, their \sysname adapter can
transparently generate and strip the checksums, as we currently do not model
data corruption.
We leave support for hardware flow control as future work.

\subsection{Inter-Simulator Message Transport}
\label{ssec:design:transport}
\sysname runs component simulators as separate processes communicating
through message passing.
Thus, efficient inter-process communication is critical for the overall
performance.
We use optimized shared memory queues with polling for efficient
message transport between simulators.
For parallel processes on separate cores, shared memory queues
enable low-latency communication with minimal
overhead~\cite{bershad:urpc,baumann:barrelfish}.
Between any pair of \emph{communicating} simulators, \sysname
establishes a bidirectional message channel consisting of a pair of
unidirectional queues in opposite directions.
During channel initialization, \sysname uses a Unix socket to provide a named
endpoint for connection setup and for communicating queue parameters and shared
memory file descriptors.

\sysname uses concurrent, circular, single-producer and consumer queues.
They comprise an array of fixed-sized, cache line aligned message
slots.
The last byte in each slot is reserved for metadata:
one bit indicating the current owner of the slot (\texttt{consumer} or
\texttt{producer}) and the rest for the message type.
As queues are single-producer and single-consumer, we store the tail pointer
\textit{locally} at the producer and the head pointer at the consumer.
Consumers poll for a message in the next slot, until the ownership
flag indicates \texttt{consumer}.
After processing the message the consumer resets the ownership flag.
Producers similarly wait for the next slot to be available, fill it,
and switch the ownership flag.

The \sysname message transport design avoids cache coherence overhead unless it
is fundamentally necessary.
Since head and tail pointers are local to consumer and producer
respectively, only accesses to shared message slots result in
coherence traffic.
Moreover, as long as a consumer does not poll in between the producer
writing a message to the corresponding slot and setting the ownership
bit, all coherence traffic carries necessary data from producer to
consumer~\cite{baumann:barrelfish}.
We include additional detail and pseudocode in \autoref{sec:appendix:shm}.

\subsection{Scaling Up with Decomposition}
\label{ssec:design:decomp}
\sysname can scale to larger simulations by adding more component simulators.
For instance, a network simulator connecting to many devices may become a
bottleneck as it needs to synchronize with all peers.
We leverage the \sysname architecture to improve scalability, by
decomposing the network simulator into multiple processes that
connect and synchronize via \sysname Ethernet interfaces.
Other simulators, such as a gem5 simulated host, can be accelerated in a similar
fashion by decomposing into connected components.
We will demonstrate the scalability benefit of our decomposition approach in
\autoref{ssec:eval:scalability}.

\subsection{Scaling Out with Proxies}
\label{ssec:design:proxy}
Running simulators in parallel on dedicated cores maximizes parallelism, but the
number of available cores in a single machine limits simulation size.
Message passing and modular simulation in \sysname enables us to scale out
simulations by partitioning components to multiple hosts and replacing message
queues between simulators on different hosts with network communication.
However, directly implementing this in individual component simulators
has two major drawbacks.
First, it increases the complexity for integration,
as each simulator adapter needs to implement an additional message
transport.
Second, it increases communication overhead in component simulators,
leaving fewer processor cycles for simulators and increasing
simulation time.
To avoid these drawbacks, we instead implement network communication
in proxies.
\sysname proxies connect to local component simulators through
existing shared memory queues and forward messages over the network to
their peer proxy which operates symmetrically.
This requires an additional processor core for the proxy on each side,
but is fully transparent to component simulators and does not increase
their communication overhead.

\subsection{Simulator Synchronization Mechanism}
\label{ssec:design:syncproto}

To ensure accurate interconnection of component simulators, we design a
synchronization mechanism that that guarantees correctness while minimizing
overhead, even when scaling to large simulations.

\subsubsection{Naive Synchronization Mechanisms do not Scale} \hfill\\
A conceptual straw-man for synchronizing components are global
barriers at each time step, keeping simulators in lockstep.
When components are connected by communication links with non-zero
latency, frequency of global barriers can be reduced by dividing
simulation time into \textit{epochs} no larger than the lowest link
latency.
Global barriers are only required at epoch boundaries, since all
cross-component events will be delivered after the end of the
current epoch~\cite{reinhardt:wwt,mohammad:distgem5,alian:pd-gem5}.
Unfortunately, epoch-based synchronization still relies on
non-scalable global barriers across all simulators, with the barrier
frequency determined by the lowest link latency in the whole
simulation, incurring substantial synchronization overhead.

\subsubsection{Scalable synchronization in \sysname} \hfill\\
We avoid global synchronization while \textit{guaranteeing accurate
simulator interconnection} by relying on properties specific to the
\sysname architecture.
\autoref{fig:sync-proto} shows pseudocode for the \sysname
synchronization protocol.

\paragraph{Enforcing message processing times is sufficient.}
In \sysname, all communication between simulators is explicit through
message passing along statically created point-to-point channels.
Thus, the only requirement for accurate simulation is that
\emph{messages are processed at the correct
time}~\cite{chandy:distsim,bryant:distsim}.
Additional synchronization does not affect the simulation, as
simulators cannot otherwise observe or influence each other.
To enforce this guarantee, senders tag messages with the time when
the receiver must process the message.
For determinism, simulators with multiple peers must order messages with
identical timestamps consistently.

\paragraph{Pairwise synchronization is sufficient.}
All \sysname message passing channels are point-to-point and
statically determined by the simulation structure.
This is where we differ from most prior synchronization schemes:
they do not assume a known topology and thus require global
synchronization.
\sysname only needs to implement pairwise synchronization, between each
simulator and its a priori known peers~\cite{bryant:distsim}.

\paragraph{Per-channel message timestamps are monotonic.}
Our message queues deliver messages strictly in order.
Since each \sysname connection between two simulators incurs a
propagation latency $\Delta_i > 0$, a message sent at time
$T$ over interface $i$ arrives at $T + \Delta_i$.
Assuming simulator clocks advance monotonically,
message timestamps on each channel are thus monotonic.

\paragraph{Message timestamps ensure correctness.}
A corollary of monotonic timestamps is that
a message with timestamp $t$ is an implicit promise that no
messages with timestamps $<t$ will arrive on that channel later.
Therefore, once a simulator receives messages with timestamps
$\ge T$ from \emph{all} its peers, it can safely advance
its clock to $T$ without more coordination.

\paragraph{Ensuring liveness with sync messages.}
The above conditions ensure accuracy, but do not guarantee liveness.
Simulations can only make progress when every channel carries at
least one message in each direction in every $\Delta_i$ time
interval~\cite{chandy:distsim,bryant:distsim}.
To ensure progress, we introduce \texttt{SYNC} messages that
simulators send if they have not sent any messages for
$\delta_i \le \Delta_i$ time units.
\texttt{SYNC} messages allow connected peers to advance their clocks in the
absence of data messages.
In our simulations we set $\delta_i = \Delta_i$;
lower values of $\delta_i$ are valid, but we have not found
configurations where the benefit of more frequent clock advances
outweighed the cost of sending and processing additional
\texttt{SYNC} messages.

\paragraph{Link latency provides synchronization slack.}
Non-zero link latencies further reduce synchronization overhead, since
not even peer simulators need to execute in lockstep.
Specifically, a message sent at $T$ allows its peer to advance to $T +
\Delta_i$.
At that point, the peer's clock is guaranteed to lay between $T -
\Delta_i$ (otherwise the local clock would not be at $T$) and
$T + \Delta_i$.
Different channels in a \sysname configuration can use different
$\Delta_i$ values.
While synchronized simulations are fundamentally only as fast as the
slowest component, this slack improves efficiency by absorbing small
transient variation in simulation speed, without immediately blocking
all simulators.

\begin{figure}%
  \begin{minipage}{\linewidth}
  \begin{algorithmic}[0]%
    \Procedure{Init}{}
      \For{if \textbf{in} interfaces}
        \State \Call{SyncTimer}{if}
        \State msg $\gets$ \Call{PollMsg}{if}
        \State \Call{Reschedule}{msg.timestamp, \textsc{RxTimer}, msg, if}
      \EndFor
    \EndProcedure
    \Procedure{SyncTimer}{if}
      \State msg $\gets$ \Call{AllocMsg}{if}
      \State msg.type $\gets$ \texttt{SYNC}
      \State \Call{SendMsg}{msg}
    \EndProcedure
    \Procedure{RxTimer}{msg, if}
      \If{msg.type $\ne$ \texttt{SYNC}}
        \State \Call{ProcessMsg}{msg}
      \EndIf
      \State msg $\gets$ \Call{PollMsg}{if}
      \State \Call{Reschedule}{msg.timestamp, \textsc{RxTimer}, msg, if}
    \EndProcedure
    \Procedure{SendMsg}{msg, if}
      \State msg.timestamp $\gets T + \Delta_{\textrm{if}}$
      \State \Call{EnqueueMsg}{msg, if}
      \State \Call{Reschedule}{$T + \delta_{\textrm{if}}$, \textsc{SyncTimer}}
    \EndProcedure
  \end{algorithmic}%
\end{minipage}%
\caption{\sysname synchronization protocol pseudocode for a
    discrete event-based simulator.
    \textsc{Reschedule} schedules a callback for the specified time,
    cancelling earlier instances.
    \textsc{ProcessMsg} and \textsc{SendMsg} interface with the
    upper layer PCI or Network protocol. $\Delta_\textrm{if}$ is the
    link latency and $\delta_\textrm{if}$ the synchronization
    interval.}%
\label{fig:sync-proto}%
\Description{This figure shows pseudo-code but the text is included as
  in the PDF and as such should be accessible.}%
\end{figure}

\section{Implementation}
\label{sec:impl}

\sysname is implemented in 4206 of C/C++ and 2102 lines of Python for
core functionality, 5348 lines for adapters in existing simulators,
and 4556 lines for new simulators we built (details in
\autoref{ssec:apppendix:implcode}).

\subsection{Core \sysname Components}
\label{ssec:impl:library}
\paragraph{Libraries.}
To reduce integration effort for simulators, we develop a common
library that implements the \sysname messaging interfaces, and
helper functions to parse and generate synchronization messages.
We also implement a helper library with common C++ components for
behavioral NIC simulators (\texttt{nicbm}) that we use for
our NIC simulators below.

\paragraph{Proxies.}
To scale out \sysname simulations,
we have implemented two proxies, one uses TCP sockets for network
communication and the other one uses RDMA.
Both implement adaptive batching by forwarding multiple messages at once if more
than one is available in the queue.
The RDMA proxy minimizes communication latency and CPU overhead by directly
writing messages to remote queues.

\paragraph{Orchestration.}
Configuring and running \sysname simulations is a challenge due to the multitude
of interconnected components involved.
We streamline simulation setups with our orchestration framework.
Users can assemble complete simulations in compact python scripts, and the
framework is responsible for running individual components (details in
\autoref{ssec:appendix:orchestration}).

\subsection{Host Simulation}
\label{ssec:impl:host}
We have integrated two host simulators, gem5 and QEMU, that are capable of
running unmodified operating systems and applications.
For both, we implement the \sysname adapter as a regular PCIe device
within the simulator's device abstractions.

\paragraph{gem5.}
gem5 is a flexible full system simulation with configurable level of
detail for memory and CPU.
We use version v20.0.0.1, extend it with a patch for Intel DDIO
support~\cite{gem5_ddio}, and implement support for the functional and timing
memory protocols.
The functional protocol is blocking, i.e., it expects device accesses and
DMA to synchronously return results, and does not model timing.
The timing protocol models accesses as asynchronous request and
response messages.
To reduce simulation time, we can configure gem5 to boot up with a fast
functional CPU, and then switch to a detailed synchronized CPU.
We also implement an Ethernet adapter to connect the built-in NICs in gem5 to
\sysname for comparison.

\paragraph{QEMU.}
We use QEMU version 5.1.92 with KVM CPU acceleration for fast
functional simulation.
We also implement support for synchronized simulation with instruction counting
(\texttt{icount}), in which QEMU controls the rate of instruction execution
relative to a virtual clock.
The key challenge is modelling MMIO delays, as QEMU's device interface does not
model timing and expects accesses to return immediately.
We work around this by aborting execution of the instruction from the
MMIO handler and stopping the virtual CPU, only re-activating it when
the \sysname PCIe completion event arrives.
QEMU will then re-try the instruction.
Unfortunately we have found that this QEMU version is no longer fully
deterministic even with instruction counting.

\subsection{NIC Simulation}
We integrate three NIC simulators, a detailed hardware RTL model, and
two less detailed but faster behavioral simulators.

\paragraph{Corundum RTL.}
To demonstrate realistic RTL device simulation, we use the unmodified Verilog
implementation of the open source Corundum FPGA NIC~\cite{forencich:corundum}.
We use Verilator~\cite{software:verilator} to simulate the \texttt{interface}
module implementing Corundum's data path, including RX, TX, descriptor queues,
checksums, and scheduling.
As Verilator cannot simulate vendor IP Corundum uses for PCIe, DMA, and
Ethernet, we implement them directly in the C++ testbench.

\paragraph{Corundum behavioral.}
To enable a fair comparison with other simulators, we also
implement a fast behavioral simulator for Corundum in C++.
Both Corundum simulators are fully compatible with the unmodified
upstream Linux driver~\cite{corundum_code}.

\paragraph{Intel i40e behavioral.}
Many recent network systems require a modern NIC compatible with
Linux or kernel-bypass frameworks such as DPDK~\cite{software:dpdk}.
We implement a behavioral simulator for the common \texttt{i40e} Intel
40G X710 NIC.
This simulator is compatible with unmodified drivers, and it implements
important NIC features such as multiple descriptor queues, TCP and IP checksum
offload, receive-side scaling, large segment offload, interrupt moderation, and
support for MSI and MSI-X.

\subsection{Network Simulation}

\paragraph{ns-3 and OMNeT++.}
To integrate with ns-3.31, we implement a \sysname Ethernet adapter
class extending \nd, the ns-3 base abstraction for host network
interfaces.
When receiving packets from our Ethernet interface, the adapter pushes them to
the connected network channel, and vice-versa.
The adapter also implements our synchronization protocol
(\autoref{fig:sync-proto}).
We integrate OMNeT++ with INET~\cite{software:omnetinet} analogously.

\paragraph{Ethernet switch.}
We also implement a fast simulator for a basic Ethernet switch.
In the simulation loop, the switch polls packets from each port,
performs MAC learning, switches each packet to the corresponding
egress port(s) according to the MAC table, and sends synchronization
messages as necessary.

\paragraph{Tofino.}
We integrate the Tofino~\cite{product:intel:tofino} simulator provided by
Intel~\cite{software:p4studio}, as the most popular programmable switch.
This simulator includes a cycle accurate model of the switch pipeline
and an approximate model for queuing.
The simulator is closed source, communicates through Linux Kernel
virtual Ethernet interfaces (\texttt{veth}), and only allows minimal
control over timing.
To implement a synchronized adapter, we parse the output log of the
simulator and generate packet timestamps accordingly.

\paragraph{Menshen RTL.}
Finally, we integrate the Verilog implementation of the Menshen RMT
pipeline~\cite{wang:menshen} using Verilator and the C++ Ethernet MAC
adapter we implemented for Corundum.

\subsection{Limitations}
\paragraph{Incompatible simulation models.}
We do not support the gem5 atomic memory protocol where memory
operations, including DMA and MMIO, are implemented as synchronous
function calls that return how long the operation should take.
This is incompatible with \sysname's asynchronous PCIe interface.
For example, while the \sysname PCIe adapter is waiting for an MMIO
completion message, no other events, such as incoming DMA requests can
be scheduled and executed.

\paragraph{Single-core hosts.}
Both gem5 and synchronized QEMU simulate multiple cores sequentially, resulting
in a super-linear increase in simulation time.
As host simulator internals are orthogonal, we pragmatically opt to
restrict our evaluation to single-core hosts.
The scalable x86 simulators we
found~\cite{miller:graphite,fu:prime,sanchez:zsim} only simulate
applications and cannot run operating systems, precluding end-to-end
simulation.
As future work, we envision applying our techniques to scale out existing full
system simulators, as modern multi-cores are essentially networked
systems~\cite{baumann:osdistsys} with message latencies.
\section{Evaluation}
\label{sec:eval}
We now evaluate if \sysname meets our design goals
(\autoref{ssec:approach:goals}):

\begin{itemize}
  \item Can \sysname \emph{modularly} combine simulators into
    \emph{end-to-end} simulations?
    How do these simulations perform?
    (\autoref{ssec:eval:compsim})
    \item How \textit{efficient} is the \sysname synchronization mechanism?
        How does the overhead compare to prior approaches?
        (\autoref{ssec:eval:syncproto})
  \item Can \sysname enable \emph{faster} simulations by breaking down
    large simulators into smaller, parallel simulators?
    (\autoref{ssec:eval:decomp})
  \item How do larger \sysname simulations \emph{scale} on a single
    physical host and distributed across multiple physical hosts?
    (\autoref{ssec:eval:scalability})
  \item Does \sysname \emph{accurately} combine simulators?
    (\autoref{ssec:eval:accurate})
  \item Are \sysname simulations \emph{deterministic}?
    (\autoref{ssec:eval:deterministic})
\end{itemize}

\subsection{Experimental Setup}
Unless otherwise stated we use the following setup:
We run simulations on physical hosts with two 22-core Intel Xeon Gold 6152
processors at 2.10\,GHz with 187\,GB of memory, hyper-threading disabled, and
100\,Gbps Mellanox ConnectX-5 NICs.%
\footnote{The testbed  only affects simulation time and unsynchronized experiments.}
All simulated hosts have a single core and 8\,GB of memory, and each runs Ubuntu
18.04 with kernel 5.4.46 where we disabled unneeded features and drivers to
reduce boot time.
All device drivers and applications are unmodified.
For synchronized QEMU we set a clock frequency of 4GHz.
For gem5, we use \texttt{DDR4\_2400\_16x4} memory and the \texttt{TimingSimple}
CPU model, which simulates an in-order CPU with the timing memory protocol, and
configure cache sizes and latencies to match those of the testbed.
We set gem5 parameters (\eg, in-order CPU clock frequency of
8\,GHz~\footnote{Gem5 also supports an out-of-order CPU, but with
$4-6\times$ higher simulation time, so we use the TmingSimple CPU as a
compromise.}) to achieve the same effective instruction execution
performance as a representative physical testbed~\cite{kaufmann:tas},
for a Linux networking benchmark at 1.3\,cycles/inst = 0.43\,ns/inst.
Further, we set the PCIe latency, Ethernet link latency and
synchronization interval all to 500\,ns, network bandwidth to
10\,Gbps, and frequency for the Corundum RTL model to 250\,MHz.

\subsection{\sysname is Modular}
\label{ssec:eval:compsim}
\paragraph{Navigating speed-accuracy trade-offs.}
We start by evaluating modular combinations of component simulators in
\sysname.
As a workload, we use the \texttt{netperf} TCP benchmark to run a 10s
throughput test (\texttt{TCP\_STREAM}) followed by a 10s latency test
(\texttt{TCP\_RR}) between two simulated hosts.
We focus on four configurations for common systems research
use-cases: debugging and performance evaluation of hardware and
software prototypes.
Debugging HW \& SW is most productive when fast and interactive,
while accurate performance is not the primary concern.
Here we combine QEMU with KVM for fast host simulation, our fast
switch model, and either the \texttt{i40e} NIC for SW testing or
Verilator with Corundum as a HW example.
Performance evaluation on the other hand requires accurate results, but it can
tolerate longer simulation times.
We use a detailed gem5 host simulator and ns-3 for SW performance evaluation,
while choosing a less detailed but time-synchronized QEMU simulator for
benchmarking our HW prototype.

\begin{table}%
\centering%
\begin{tabular}{lrrrr}%
    \toprule
    \textbf{Use-case} & \multicolumn{2}{c}{netperf} &
        \multicolumn{1}{c}{Sim.} \\
    Simulator Combination & T'put & Latency &
        \multicolumn{1}{c}{Time} \\
    \midrule

    \textbf{SW debugging} & 4.37\,G & 71\,$\mu$s & 00:00:32 \\
    \multicolumn{4}{l}{QEMU-kvm + behavioral i40e NIC + behavioral
      switch} \\[.3em]

    \textbf{SW perf. evaluation*} & 8.92\,G & 20\,$\mu$s & 12:49:46  \\
    \multicolumn{4}{l}{gem5 + behavioral i40e NIC + ns-3} \\[.3em]

    \textbf{HW debugging} & 81\,M & 3.4\,ms & 00:00:31 \\
    \multicolumn{4}{l}{QEMU-kvm + Corundum Verilog + behavioral
      switch} \\[.3em]

    \textbf{HW perf. evaluation*} & 6.55\,G & 32\,$\mu$s & 04:13:10 \\
    \multicolumn{4}{l}{QEMU-timing + Corundum Verilog + behavioral
      switch}\\
    \bottomrule\\
\end{tabular}%
\caption{\sysname configurations for different use-cases, with
    measured simulation time and application performance.
    Configurations with * are synchronized and deterministic, while
    the others are unsynchronized emulation.}%
\label{tab:modcombo}%
\vspace{-5mm}%
\end{table}

Our results in \autoref{tab:modcombo} confirm the expected trade-off between
simulation time and simulator detail: simulation times range from 31s to 18
hours.
The results show that, \sysname can effectively help navigate this trade-off by
only using detailed simulators when details matter for the use-case.
Even combining fast QEMU-kvm with an unsynchronized RTL simulation is
fast enough (31s) to test and debug the full system.
Modularity also allows us to late bind simulator choices, \eg if we
later realize that QEMU-timing is not sufficiently accurate, we can
replace it with gem5 without additional changes.

\paragraph{All combinations are functional.}
Besides these four configurations, we also evaluated the full cross-product of
simulator choices (\autoref{sec:impl}) and confirm \sysname supports all
combinations (subset of performance results in
\autoref{ssec:appendix:simcombos}).

\paragraph{\sysname interfaces are general.}
\sysname interfaces are generic and serve as narrow waists between
simulators.
To further demonstrate its generality, we extracted gem5's
\texttt{e1000} Intel NIC model, adapted it to \sysname's PCIe
interface without other modifications, and verified that it is
compatible with gem5 and QEMU.
To show that \sysname's PCIe interface generalizes beyond NICs, we
have adapted FEMU~\cite{li:femu}'s NVMe SSD model from their QEMU
fork into a separate simulator.
This simulator also works in combination with QEMU and gem5.

\subsection{\sysname is Fast}
We now show \sysname does not significantly slow down simulators
through synchronization, and can even speed up simulations through
decomposition into parallel components.

\subsubsection{Synchronization}
\label{ssec:eval:syncproto}

\paragraph{Overhead.}
We measure synchronization overhead by comparing simulation time
for gem5 standalone and in \sysname.
The experiment does not use the network, but for synchronization, we
connect the gem5 to \texttt{i40e} NIC in \sysname and to our switch.
We first compare a low-event workload in gem5: executing
\texttt{sleep 10}.
The simulation takes 2.25\,min standalone and 2.91\,min in \sysname, a
30\% overhead.
This is the worst case -- gem5 is almost exclusively handling
\sysname synchronization events (every 500\,ns), as the CPU is mostly
halted.
For a high-event workload we use \texttt{dd} to read from
\texttt{/dev/urandom} to keep the CPU busy.
This simulation takes 100.26\,min standalone and 101.06\,min in
\sysname, a mere 0.8\% overhead.
\emph{\sysname incurs manageable synchronization overhead, and
does not significantly slow down already slow simulations.}

\paragraph{Comparison to dist-gem5.}
Next, we compare to dist-gem5~\cite{mohammad:distgem5} which
interconnects multiple gem5 instances and employs conventional
epoch-based global synchronization over TCP.
We configure 2 to 32 instances of gem5 that communicate pairwise using
iperf, through the \texttt{e1000} NIC in gem5 and a single switch.
For \sysname we use our gem5 Ethernet adapter to connect to our
switch model.
Our simulation time measurements in \autoref{fig:dist-gem} show that
\emph{\sysname is more efficient than dist-gem5, especially with
increasing scale}.
\sysname reduces simulation time by 27\% for 2 hosts, and by 74\% for 32 hosts.

\paragraph{Sensitivity to link latency.}
\sysname synchronization overhead is linked with the configured link
latency, which places a lower bound on sync message frequency.
We measure how link latency affects synchronization overhead, with
a pair of gem5 hosts running \texttt{netperf} for 1\,s of throughput
and latency measurements each, connected to \texttt{i40e} NICs and a
shared switch.
We vary the configured PCIe latency and sync interval, and report our
results in \autoref{fig:pcilat}.
While synchronization time does increase, \emph{lowering the link
latency by three orders of magnitude (from 1\,$\mu$s to 1\,ns) only
increases simulation time by 59\%}, demonstrating that \sysname can
effectively parallelize simulations across low-latency interconnects.

\begin{figure*}%
\centering%
\begin{minipage}{0.3\textwidth}%
\centering%
\begin{tikzpicture}[gnuplot]
\tikzset{every node/.append style={font={\fontsize{8.0pt}{9.6pt}\selectfont}}}
\path (0.000,0.000) rectangle (5.334,4.216);
\gpcolor{color=gp lt color border}
\gpsetlinetype{gp lt border}
\gpsetdashtype{gp dt solid}
\gpsetlinewidth{1.00}
\draw[gp path] (1.201,0.688)--(1.381,0.688);
\draw[gp path] (4.892,0.688)--(4.712,0.688);
\node[gp node right] at (1.201,0.688) {$0$};
\draw[gp path] (1.201,1.488)--(1.381,1.488);
\draw[gp path] (4.892,1.488)--(4.712,1.488);
\node[gp node right] at (1.201,1.488) {$500$};
\draw[gp path] (1.201,2.288)--(1.381,2.288);
\draw[gp path] (4.892,2.288)--(4.712,2.288);
\node[gp node right] at (1.201,2.288) {$1000$};
\draw[gp path] (1.201,3.089)--(1.381,3.089);
\draw[gp path] (4.892,3.089)--(4.712,3.089);
\node[gp node right] at (1.201,3.089) {$1500$};
\draw[gp path] (1.201,3.889)--(1.381,3.889);
\draw[gp path] (4.892,3.889)--(4.712,3.889);
\node[gp node right] at (1.201,3.889) {$2000$};
\node[gp node center] at (1.634,0.515) {2};
\node[gp node center] at (2.576,0.515) {8};
\node[gp node center] at (3.517,0.515) {16};
\node[gp node center] at (4.459,0.515) {32};
\draw[gp path] (1.201,3.969)--(1.201,0.688)--(4.892,0.688)--(4.892,3.969)--cycle;
\node[gp node center,rotate=-270] at (0.232,2.328) {Simulation Time [Min.]};
\node[gp node center] at (3.046,0.171) {Number of Simulated Hosts};
\node[gp node right] at (2.671,3.641) {dist-gem5};
\gpfill{rgb color={0.580,0.000,0.827}} (2.818,3.580)--(3.586,3.580)--(3.586,3.703)--(2.818,3.703)--cycle;
\gpcolor{rgb color={0.580,0.000,0.827}}
\draw[gp path] (2.818,3.580)--(3.586,3.580)--(3.586,3.702)--(2.818,3.702)--cycle;
\gpfill{rgb color={0.580,0.000,0.827}} (1.342,0.688)--(1.626,0.688)--(1.626,1.467)--(1.342,1.467)--cycle;
\draw[gp path] (1.342,0.688)--(1.342,1.466)--(1.625,1.466)--(1.625,0.688)--cycle;
\gpfill{rgb color={0.580,0.000,0.827}} (2.284,0.688)--(2.567,0.688)--(2.567,1.789)--(2.284,1.789)--cycle;
\draw[gp path] (2.284,0.688)--(2.284,1.788)--(2.566,1.788)--(2.566,0.688)--cycle;
\gpfill{rgb color={0.580,0.000,0.827}} (3.225,0.688)--(3.509,0.688)--(3.509,2.134)--(3.225,2.134)--cycle;
\draw[gp path] (3.225,0.688)--(3.225,2.133)--(3.508,2.133)--(3.508,0.688)--cycle;
\gpfill{rgb color={0.580,0.000,0.827}} (4.167,0.688)--(4.450,0.688)--(4.450,3.554)--(4.167,3.554)--cycle;
\draw[gp path] (4.167,0.688)--(4.167,3.553)--(4.449,3.553)--(4.449,0.688)--cycle;
\gpcolor{color=gp lt color border}
\node[gp node right] at (2.671,3.346) {\sysname};
\gpfill{rgb color={0.000,0.620,0.451}} (2.818,3.285)--(3.586,3.285)--(3.586,3.408)--(2.818,3.408)--cycle;
\gpcolor{rgb color={0.000,0.620,0.451}}
\draw[gp path] (2.818,3.285)--(3.586,3.285)--(3.586,3.407)--(2.818,3.407)--cycle;
\gpfill{rgb color={0.000,0.620,0.451}} (1.644,0.688)--(1.927,0.688)--(1.927,1.249)--(1.644,1.249)--cycle;
\draw[gp path] (1.644,0.688)--(1.644,1.248)--(1.926,1.248)--(1.926,0.688)--cycle;
\gpfill{rgb color={0.000,0.620,0.451}} (2.585,0.688)--(2.869,0.688)--(2.869,1.265)--(2.585,1.265)--cycle;
\draw[gp path] (2.585,0.688)--(2.585,1.264)--(2.868,1.264)--(2.868,0.688)--cycle;
\gpfill{rgb color={0.000,0.620,0.451}} (3.527,0.688)--(3.810,0.688)--(3.810,1.291)--(3.527,1.291)--cycle;
\draw[gp path] (3.527,0.688)--(3.527,1.290)--(3.809,1.290)--(3.809,0.688)--cycle;
\gpfill{rgb color={0.000,0.620,0.451}} (4.468,0.688)--(4.752,0.688)--(4.752,1.430)--(4.468,1.430)--cycle;
\draw[gp path] (4.468,0.688)--(4.468,1.429)--(4.751,1.429)--(4.751,0.688)--cycle;
\gpcolor{color=gp lt color border}
\node[gp node center] at (1.634,1.658) {-27\%};
\node[gp node center] at (2.576,1.980) {-47\%};
\node[gp node center] at (3.517,2.325) {-58\%};
\node[gp node center] at (4.459,3.745) {-74\%};
\draw[gp path] (1.201,3.969)--(1.201,0.688)--(4.892,0.688)--(4.892,3.969)--cycle;
\gpdefrectangularnode{gp plot 1}{\pgfpoint{1.201cm}{0.688cm}}{\pgfpoint{4.892cm}{3.969cm}}
\end{tikzpicture}
\caption{dist-gem5 vs. \sysname.}%
\label{fig:dist-gem}%
\end{minipage}%
\hfill%
\hfill%
\begin{minipage}{0.3\textwidth}%
\centering%
\begin{tikzpicture}[gnuplot]
\tikzset{every node/.append style={font={\fontsize{8.0pt}{9.6pt}\selectfont}}}
\path (0.000,0.000) rectangle (5.334,4.216);
\gpcolor{color=gp lt color border}
\gpsetlinetype{gp lt border}
\gpsetdashtype{gp dt solid}
\gpsetlinewidth{1.00}
\draw[gp path] (1.054,0.688)--(1.234,0.688);
\draw[gp path] (4.892,0.688)--(4.712,0.688);
\node[gp node right] at (1.054,0.688) {$0$};
\draw[gp path] (1.054,1.469)--(1.234,1.469);
\draw[gp path] (4.892,1.469)--(4.712,1.469);
\node[gp node right] at (1.054,1.469) {$50$};
\draw[gp path] (1.054,2.250)--(1.234,2.250);
\draw[gp path] (4.892,2.250)--(4.712,2.250);
\node[gp node right] at (1.054,2.250) {$100$};
\draw[gp path] (1.054,3.032)--(1.234,3.032);
\draw[gp path] (4.892,3.032)--(4.712,3.032);
\node[gp node right] at (1.054,3.032) {$150$};
\draw[gp path] (1.054,3.813)--(1.234,3.813);
\draw[gp path] (4.892,3.813)--(4.712,3.813);
\node[gp node right] at (1.054,3.813) {$200$};
\draw[gp path] (1.709,0.688)--(1.709,0.868);
\draw[gp path] (1.709,3.969)--(1.709,3.789);
\node[gp node center] at (1.709,0.515) {$5$};
\draw[gp path] (2.645,0.688)--(2.645,0.868);
\draw[gp path] (2.645,3.969)--(2.645,3.789);
\node[gp node center] at (2.645,0.515) {$10$};
\draw[gp path] (3.581,0.688)--(3.581,0.868);
\draw[gp path] (3.581,3.969)--(3.581,3.789);
\node[gp node center] at (3.581,0.515) {$15$};
\draw[gp path] (4.518,0.688)--(4.518,0.868);
\draw[gp path] (4.518,3.969)--(4.518,3.789);
\node[gp node center] at (4.518,0.515) {$20$};
\draw[gp path] (1.054,3.969)--(1.054,0.688)--(4.892,0.688)--(4.892,3.969)--cycle;
\node[gp node center,rotate=-270] at (0.232,2.328) {Simulation Time [Min.]};
\node[gp node center] at (2.973,0.171) {Number of Simulated Hosts};
\gpcolor{rgb color={0.580,0.000,0.827}}
\draw[gp path] (1.148,2.845)--(1.709,2.881)--(2.645,3.345)--(3.581,3.271)--(4.705,3.901);
\gpsetpointsize{4.00}
\gp3point{gp mark 1}{}{(1.148,2.845)}
\gp3point{gp mark 1}{}{(1.709,2.881)}
\gp3point{gp mark 1}{}{(2.645,3.345)}
\gp3point{gp mark 1}{}{(3.581,3.271)}
\gp3point{gp mark 1}{}{(4.705,3.901)}
\gpcolor{color=gp lt color border}
\draw[gp path] (1.054,3.969)--(1.054,0.688)--(4.892,0.688)--(4.892,3.969)--cycle;
\gpdefrectangularnode{gp plot 1}{\pgfpoint{1.054cm}{0.688cm}}{\pgfpoint{4.892cm}{3.969cm}}
\end{tikzpicture}
\caption{\sysname local scalability.}%
\label{fig:machines-scale}%
\end{minipage}%
\hfill%
\begin{minipage}{0.3\textwidth}%
\centering%
\begin{tikzpicture}[gnuplot]
\tikzset{every node/.append style={font={\fontsize{8.0pt}{9.6pt}\selectfont}}}
\path (0.000,0.000) rectangle (5.334,4.216);
\gpcolor{color=gp lt color border}
\gpsetlinetype{gp lt border}
\gpsetdashtype{gp dt solid}
\gpsetlinewidth{1.00}
\draw[gp path] (1.201,0.688)--(1.381,0.688);
\draw[gp path] (4.892,0.688)--(4.712,0.688);
\node[gp node right] at (1.201,0.688) {$0$};
\draw[gp path] (1.201,1.285)--(1.381,1.285);
\draw[gp path] (4.892,1.285)--(4.712,1.285);
\node[gp node right] at (1.201,1.285) {$200$};
\draw[gp path] (1.201,1.881)--(1.381,1.881);
\draw[gp path] (4.892,1.881)--(4.712,1.881);
\node[gp node right] at (1.201,1.881) {$400$};
\draw[gp path] (1.201,2.478)--(1.381,2.478);
\draw[gp path] (4.892,2.478)--(4.712,2.478);
\node[gp node right] at (1.201,2.478) {$600$};
\draw[gp path] (1.201,3.074)--(1.381,3.074);
\draw[gp path] (4.892,3.074)--(4.712,3.074);
\node[gp node right] at (1.201,3.074) {$800$};
\draw[gp path] (1.201,3.671)--(1.381,3.671);
\draw[gp path] (4.892,3.671)--(4.712,3.671);
\node[gp node right] at (1.201,3.671) {$1000$};
\draw[gp path] (1.201,0.688)--(1.201,0.868);
\node[gp node center] at (1.201,0.515) {$0$};
\draw[gp path] (1.911,0.688)--(1.911,0.868);
\node[gp node center] at (1.911,0.515) {$200$};
\draw[gp path] (2.621,0.688)--(2.621,0.868);
\node[gp node center] at (2.621,0.515) {$400$};
\draw[gp path] (3.330,0.688)--(3.330,0.868);
\node[gp node center] at (3.330,0.515) {$600$};
\draw[gp path] (4.040,0.688)--(4.040,0.868);
\node[gp node center] at (4.040,0.515) {$800$};
\draw[gp path] (4.750,0.688)--(4.750,0.868);
\node[gp node center] at (4.750,0.515) {$1000$};
\draw[gp path] (1.201,3.969)--(1.201,0.688)--(4.892,0.688)--(4.892,3.969)--cycle;
\node[gp node center,rotate=-270] at (0.232,2.328) {Simulation Time [Min.]};
\node[gp node center] at (3.046,0.171) {Number of Simulated Hosts};
\node[gp node right] at (3.683,2.475) {gem5};
\gpcolor{rgb color={0.580,0.000,0.827}}
\draw[gp path] (3.830,2.475)--(4.598,2.475);
\draw[gp path] (1.343,3.462)--(4.750,3.845);
\gpsetpointsize{4.00}
\gp3point{gp mark 1}{}{(1.343,3.462)}
\gp3point{gp mark 1}{}{(4.750,3.845)}
\gp3point{gp mark 1}{}{(4.214,2.475)}
\gpcolor{color=gp lt color border}
\node[gp node right] at (3.683,2.180) {QEMU-timing};
\gpcolor{rgb color={0.000,0.620,0.451}}
\draw[gp path] (3.830,2.180)--(4.598,2.180);
\draw[gp path] (1.343,1.082)--(1.911,1.243)--(2.621,1.386)--(3.330,1.493)--(4.750,1.690);
\gp3point{gp mark 2}{}{(1.343,1.082)}
\gp3point{gp mark 2}{}{(1.911,1.243)}
\gp3point{gp mark 2}{}{(2.621,1.386)}
\gp3point{gp mark 2}{}{(3.330,1.493)}
\gp3point{gp mark 2}{}{(4.750,1.690)}
\gp3point{gp mark 2}{}{(4.214,2.180)}
\gpcolor{color=gp lt color border}
\draw[gp path] (1.201,3.969)--(1.201,0.688)--(4.892,0.688)--(4.892,3.969)--cycle;
\gpdefrectangularnode{gp plot 1}{\pgfpoint{1.201cm}{0.688cm}}{\pgfpoint{4.892cm}{3.969cm}}
\end{tikzpicture}
\caption{Distributed scalability.}%
\label{fig:large-scale}%
\end{minipage}%
\Description{Three sub-figures (6, 7, and 8) with graphs showing simulation
  time on the y-axis.

  The first graph (figure 6) shows simulation time of
  dist-gem5 compared to \sysname for varying number of simulated hosts on the
  x-axis from 2 to 32. dist-gem5 is slower across the board and shows
  exponentially increasing simulation time, while for \sysname it only slowly
  increases. For two hosts \sysname is 27\% faster at around 360 min, and 74\%
  faster for 32 at close to 450 min.

  The second figure shows a line graph for gem5 in \sysname as the number of
  hosts increases from 2 to 21. Bettween 2 and 5 the line is flat at 140
  minutes, increasing to about 170 for 10 and 15, finally increasing to around
  210 for 21 simulated hosts.

  The third figure is another line graph, comparing gem5 and QEMU-timing for a
  distributed experiment. For QEMU-timing simulation time increases
  approximately linearly from 130 min with 40 hosts to about 330 with
  1000 hosts. With gem-5 it increases from 930 min with 40 hosts to
  about 1060 with 1000 hosts.}%
\end{figure*}

\subsubsection{Decomposition for Parallelism}
\label{ssec:eval:decomp}
\paragraph{Extracting NIC from gem5.}
When connecting synchronized simulators, the best \sysname can achieve
is to not slow them down beyond the slowest component simulator.
However, \sysname enables developers to decompose monolithic
simulators into connected components (\autoref{ssec:eval:compsim})
running in parallel, thereby accelerating simulation.
We evaluate this by comparing two gem5 configurations in \sysname:
first, gem5 with the built-in \texttt{e1000} NIC connected via our
Ethernet adapter, and second, gem5 connected to our \texttt{i40e} NIC
model through the PCIe interface.
In both cases we run a pair of hosts connected to our switch model.
The first configuration takes 350 minutes, while the second only takes 138
minutes: \emph{Parallelism from the external NIC simulator reduces simulation
time by 60\%}.

\paragraph{Network simulator as scalability bottleneck.}
Network simulators are potential scalability bottlenecks in \sysname,
as they often connect many NICs, while hosts and NICs typically only connect
one and two peers, respectively.
To demonstrate this bottleneck, we develop a packet generator as a dummy NIC
that implements the \sysname Ethernet interface and the synchronization
mechanism.
The dummy NIC simply injects packets at a configured rate.
We now measure simulation time for 2 and 32 dummy NICs connected to
one switch for 1 second of virtual time.
First we set the packet rate to 0 (to only measure synchronization overhead) and
measure an increase from 2.6\,s to 17.6\,s of simulation time.
Next, we set the packet rate to 100\,Gbps on each NIC, and measure the
simulation time increases from 12.6\,s to 211.6\,s.
This experiment confirms that \emph{a single network simulator can become a
bottleneck for fast simulations}.
We have so far not observed this outside of this microbenchmark.

\paragraph{Parallelizing network simulation.}
To address this bottleneck in \sysname, we can decompose the network into
multiple network simulators carved up at natural boundaries (\eg switches or
groups thereof).
We demonstrate this by modifying the microbenchmark to divide the 32
hosts to 4 ``ToR'' switches, connected through a fifth ``core''
switch.
With this configuration, the simulation time for packet rate 0 is 9.6\,s down by
45\% compared to the single switch setup, and 96.8\,s at 100\,Gbps packet rate,
a 53\% reduction.
\emph{Decomposing network simulators, therefore, can effectively reduce
simulation time at scale.}

\subsection{\sysname is Scalable}
\label{ssec:eval:scalability}
We now evaluate scalability for local and distributed simulations.

\subsubsection{Scaling Up}\hfill\\
First, we measure simulation time as we vary the number of simulated
gem5 hosts and \texttt{i40e} NICs connected to a single switch,
running on a single physical host.
We set up one server and a variable number of client hosts, running
the same UDP iperf benchmark.
To avoid overloading the server, we fix the aggregate throughput to
1\,Gbps.
The results in \autoref{fig:machines-scale} show the simulation time
increases with the number of clients, from 138\,min with 2
hosts to 205\,min with 21 hosts (48\% increase).

Surprisingly, the longer simulation time is \emph{not} caused by scalability
bottlenecks in \sysname synchronization.
Instead, we discovered that this increase is due to thermal
throttling of our host CPU slowing down all cores as more active.
To confirm this, we run multiple independent instances of the
1-client experiment and measure how this affects simulation time.
When running 4 independent instances of the 2-host simulations (5
cores each), using a total of 20 cores in the same NUMA node, the
simulation takes 171\,min.
This matches the runtime of the 10-host simulation above, which uses
21 cores in total.
We conclude that \emph{\sysname scales at least to the moderate
cluster sizes typical for many of our evaluations}.

\begin{figure*}%
\centering%
\begin{minipage}{0.28\textwidth}%
\centering%
\begin{tikzpicture}[gnuplot]
\tikzset{every node/.append style={font={\fontsize{8.0pt}{9.6pt}\selectfont}}}
\path (0.000,0.000) rectangle (5.334,4.216);
\gpcolor{color=gp lt color border}
\gpsetlinetype{gp lt border}
\gpsetdashtype{gp dt solid}
\gpsetlinewidth{1.00}
\draw[gp path] (1.054,0.688)--(1.234,0.688);
\draw[gp path] (4.892,0.688)--(4.712,0.688);
\node[gp node right] at (1.054,0.688) {$0$};
\draw[gp path] (1.054,1.235)--(1.234,1.235);
\draw[gp path] (4.892,1.235)--(4.712,1.235);
\node[gp node right] at (1.054,1.235) {$20$};
\draw[gp path] (1.054,1.782)--(1.234,1.782);
\draw[gp path] (4.892,1.782)--(4.712,1.782);
\node[gp node right] at (1.054,1.782) {$40$};
\draw[gp path] (1.054,2.329)--(1.234,2.329);
\draw[gp path] (4.892,2.329)--(4.712,2.329);
\node[gp node right] at (1.054,2.329) {$60$};
\draw[gp path] (1.054,2.875)--(1.234,2.875);
\draw[gp path] (4.892,2.875)--(4.712,2.875);
\node[gp node right] at (1.054,2.875) {$80$};
\draw[gp path] (1.054,3.422)--(1.234,3.422);
\draw[gp path] (4.892,3.422)--(4.712,3.422);
\node[gp node right] at (1.054,3.422) {$100$};
\draw[gp path] (1.054,3.969)--(1.234,3.969);
\draw[gp path] (4.892,3.969)--(4.712,3.969);
\node[gp node right] at (1.054,3.969) {$120$};
\draw[gp path] (1.822,0.688)--(1.822,0.868);
\node[gp node center] at (1.822,0.442) {1};
\draw[gp path] (2.589,0.688)--(2.589,0.868);
\node[gp node center] at (2.589,0.442) {10};
\draw[gp path] (3.357,0.688)--(3.357,0.868);
\node[gp node center] at (3.357,0.442) {100};
\draw[gp path] (4.124,0.688)--(4.124,0.868);
\node[gp node center] at (4.124,0.442) {1000};
\draw[gp path] (1.054,3.969)--(1.054,0.688)--(4.892,0.688)--(4.892,3.969)--cycle;
\node[gp node center,rotate=-270] at (0.232,2.328) {Simulation Time [Min.]};
\node[gp node center] at (2.973,0.171) {PCIe latency [ns]};
\gpfill{rgb color={0.416,0.318,0.639}} (1.630,0.688)--(2.015,0.688)--(2.015,3.779)--(1.630,3.779)--cycle;
\gpcolor{rgb color={0.416,0.318,0.639}}
\gpsetlinewidth{2.00}
\draw[gp path] (1.630,0.688)--(1.630,3.778)--(2.014,3.778)--(2.014,0.688)--cycle;
\gpfill{rgb color={0.416,0.318,0.639}} (2.397,0.688)--(2.782,0.688)--(2.782,3.259)--(2.397,3.259)--cycle;
\draw[gp path] (2.397,0.688)--(2.397,3.258)--(2.781,3.258)--(2.781,0.688)--cycle;
\gpfill{rgb color={0.416,0.318,0.639}} (3.165,0.688)--(3.550,0.688)--(3.550,3.314)--(3.165,3.314)--cycle;
\draw[gp path] (3.165,0.688)--(3.165,3.313)--(3.549,3.313)--(3.549,0.688)--cycle;
\gpfill{rgb color={0.416,0.318,0.639}} (3.933,0.688)--(4.317,0.688)--(4.317,2.630)--(3.933,2.630)--cycle;
\draw[gp path] (3.933,0.688)--(3.933,2.629)--(4.316,2.629)--(4.316,0.688)--cycle;
\gpcolor{color=gp lt color border}
\gpsetlinewidth{1.00}
\draw[gp path] (1.822,3.746)--(1.822,3.809);
\draw[gp path] (1.732,3.746)--(1.912,3.746);
\draw[gp path] (1.732,3.809)--(1.912,3.809);
\draw[gp path] (2.589,3.234)--(2.589,3.282);
\draw[gp path] (2.499,3.234)--(2.679,3.234);
\draw[gp path] (2.499,3.282)--(2.679,3.282);
\draw[gp path] (3.357,3.299)--(3.357,3.327);
\draw[gp path] (3.267,3.299)--(3.447,3.299);
\draw[gp path] (3.267,3.327)--(3.447,3.327);
\draw[gp path] (4.124,2.620)--(4.124,2.638);
\draw[gp path] (4.034,2.620)--(4.214,2.620);
\draw[gp path] (4.034,2.638)--(4.214,2.638);
\draw[gp path] (1.054,3.969)--(1.054,0.688)--(4.892,0.688)--(4.892,3.969)--cycle;
\gpdefrectangularnode{gp plot 1}{\pgfpoint{1.054cm}{0.688cm}}{\pgfpoint{4.892cm}{3.969cm}}
\end{tikzpicture}
\caption{Sensitivity of \sysname simulation time to link latency.}%
\label{fig:pcilat}%
\end{minipage}%
\hfill%
\begin{minipage}{0.45\textwidth}%
\centering%
\begin{tikzpicture}[gnuplot]
\tikzset{every node/.append style={font={\fontsize{8.0pt}{9.6pt}\selectfont}}}
\path (0.000,0.000) rectangle (8.458,4.216);
\gpcolor{color=gp lt color border}
\gpsetlinetype{gp lt border}
\gpsetdashtype{gp dt solid}
\gpsetlinewidth{1.00}
\draw[gp path] (1.054,0.688)--(1.234,0.688);
\draw[gp path] (8.016,0.688)--(7.836,0.688);
\node[gp node right] at (1.054,0.688) {$0$};
\draw[gp path] (1.054,1.285)--(1.234,1.285);
\draw[gp path] (8.016,1.285)--(7.836,1.285);
\node[gp node right] at (1.054,1.285) {$20$};
\draw[gp path] (1.054,1.881)--(1.234,1.881);
\draw[gp path] (8.016,1.881)--(7.836,1.881);
\node[gp node right] at (1.054,1.881) {$40$};
\draw[gp path] (1.054,2.478)--(1.234,2.478);
\draw[gp path] (8.016,2.478)--(7.836,2.478);
\node[gp node right] at (1.054,2.478) {$60$};
\draw[gp path] (1.054,3.074)--(1.234,3.074);
\draw[gp path] (8.016,3.074)--(7.836,3.074);
\node[gp node right] at (1.054,3.074) {$80$};
\draw[gp path] (1.054,3.671)--(1.234,3.671);
\draw[gp path] (8.016,3.671)--(7.836,3.671);
\node[gp node right] at (1.054,3.671) {$100$};
\draw[gp path] (1.054,0.688)--(1.054,0.868);
\draw[gp path] (1.054,3.969)--(1.054,3.789);
\node[gp node center] at (1.054,0.515) {$0$};
\draw[gp path] (1.873,0.688)--(1.873,0.868);
\draw[gp path] (1.873,3.969)--(1.873,3.789);
\node[gp node center] at (1.873,0.515) {$10$};
\draw[gp path] (2.692,0.688)--(2.692,0.868);
\draw[gp path] (2.692,3.969)--(2.692,3.789);
\node[gp node center] at (2.692,0.515) {$20$};
\draw[gp path] (3.511,0.688)--(3.511,0.868);
\draw[gp path] (3.511,3.969)--(3.511,3.789);
\node[gp node center] at (3.511,0.515) {$30$};
\draw[gp path] (4.330,0.688)--(4.330,0.868);
\draw[gp path] (4.330,3.969)--(4.330,3.789);
\node[gp node center] at (4.330,0.515) {$40$};
\draw[gp path] (5.149,0.688)--(5.149,0.868);
\draw[gp path] (5.149,3.969)--(5.149,3.789);
\node[gp node center] at (5.149,0.515) {$50$};
\draw[gp path] (5.968,0.688)--(5.968,0.868);
\draw[gp path] (5.968,3.969)--(5.968,3.789);
\node[gp node center] at (5.968,0.515) {$60$};
\draw[gp path] (6.787,0.688)--(6.787,0.868);
\draw[gp path] (6.787,3.969)--(6.787,3.789);
\node[gp node center] at (6.787,0.515) {$70$};
\draw[gp path] (7.606,0.688)--(7.606,0.868);
\draw[gp path] (7.606,3.969)--(7.606,3.789);
\node[gp node center] at (7.606,0.515) {$80$};
\draw[gp path] (1.054,3.969)--(1.054,0.688)--(8.016,0.688)--(8.016,3.969)--cycle;
\node[gp node center,rotate=-270] at (0.232,2.328) {Latency [us]};
\node[gp node center] at (4.535,0.171) {Throughput [Krequest/sec]};
\node[gp node right] at (6.807,1.605) {End-host Sequencer};
\gpcolor{rgb color={0.580,0.000,0.827}}
\draw[gp path] (6.954,1.605)--(7.722,1.605);
\draw[gp path] (2.512,2.269)--(3.908,2.239)--(6.387,2.299)--(7.344,2.418)--(7.450,2.686)%
  --(7.438,3.462)--(7.421,3.969);
\gpsetpointsize{4.00}
\gp3point{gp mark 1}{}{(2.512,2.269)}
\gp3point{gp mark 1}{}{(3.908,2.239)}
\gp3point{gp mark 1}{}{(6.387,2.299)}
\gp3point{gp mark 1}{}{(7.344,2.418)}
\gp3point{gp mark 1}{}{(7.450,2.686)}
\gp3point{gp mark 1}{}{(7.438,3.462)}
\gp3point{gp mark 1}{}{(7.338,1.605)}
\gpcolor{color=gp lt color border}
\node[gp node right] at (6.807,1.310) {Switch Sequencer};
\gpcolor{rgb color={0.000,0.620,0.451}}
\draw[gp path] (6.954,1.310)--(7.722,1.310);
\draw[gp path] (2.846,1.971)--(4.537,1.971)--(7.281,2.060)--(7.395,2.746)--(7.357,3.551)%
  --(7.343,3.969);
\gp3point{gp mark 2}{}{(2.846,1.971)}
\gp3point{gp mark 2}{}{(4.537,1.971)}
\gp3point{gp mark 2}{}{(7.281,2.060)}
\gp3point{gp mark 2}{}{(7.395,2.746)}
\gp3point{gp mark 2}{}{(7.357,3.551)}
\gp3point{gp mark 2}{}{(7.338,1.310)}
\gpcolor{color=gp lt color border}
\node[gp node right] at (6.807,1.015) {Multi-Paxos};
\gpcolor{rgb color={0.337,0.706,0.914}}
\draw[gp path] (6.954,1.015)--(7.722,1.015);
\draw[gp path] (2.165,2.806)--(2.811,3.372)--(2.815,3.969);
\gp3point{gp mark 3}{}{(2.165,2.806)}
\gp3point{gp mark 3}{}{(2.811,3.372)}
\gp3point{gp mark 3}{}{(7.338,1.015)}
\gpcolor{color=gp lt color border}
\draw[gp path] (1.054,3.969)--(1.054,0.688)--(8.016,0.688)--(8.016,3.969)--cycle;
\gpdefrectangularnode{gp plot 1}{\pgfpoint{1.054cm}{0.688cm}}{\pgfpoint{8.016cm}{3.969cm}}
\end{tikzpicture}
\caption{NOPaxos in \sysname with a Tofino switch sequencer and with a
    sequencer on a simulated host.}%
\label{fig:nopaxos-seq}%
\end{minipage}%
\hfill%
\begin{minipage}{0.2\textwidth}%
\centering%
\vspace{4mm}
\includegraphics[width=0.8\linewidth]{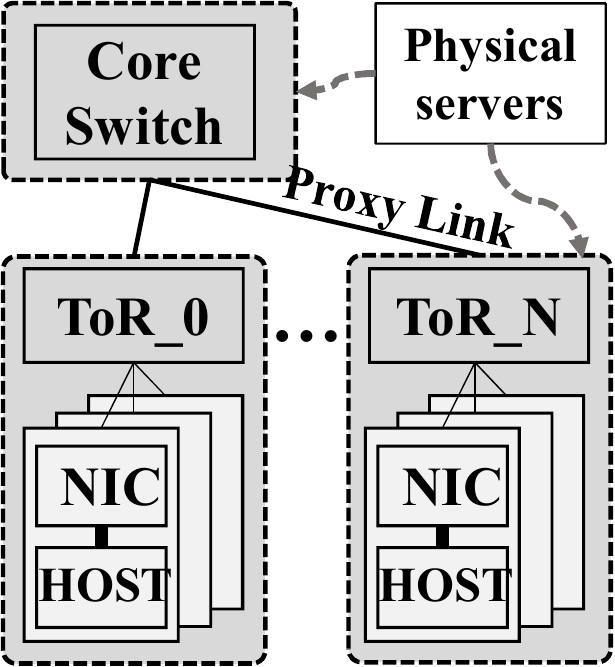}
\vspace{8mm}
\caption{\textls[-15]{Large scale simulation configuration.}}
\label{fig:largescale-config}
\end{minipage}%
\Description{Three subfigures: 9, 10, 11.

  The first subfigure (fig. 9) is a bar graph with PCIe latency on the
  x-axis and simulation time on the y-axis. As the latency increases
  from 1\,ns to 1000\,ns simulation time decreases from 117 minutes to
  about 72.

  The second subfigure (fig. 10) is a line graph comparing latency on
  the y-axis for end-host sequencer, switch-sequencer, and Multi-Paxos
  as throughput on the x-axis increases. The lowest line is the switch
  sequencer flat at around 40 microseconds starting at 20 thousand
  requests per second, until it hist saturation and rapidly increases
  at 75 thousand requests per second. The end-host sequencer
  configuration achieves a flat higher latency at around 55
  microseconds and saturates at the same point. Multi-Paxos already
  incurs 70 microseconds at 12 thousand requests, 90 microseconds at
  20 thousand requests, and then saturates.

  The third subfigure (fig. 11) is a diagram showing the large-scale
  simulation configuration, with boxes representing groups of
  simulators running on separate physical server. Each server contains
  a group of simulators and connects to one central server through
  proxy links. One server contains just one core switch. The remaining
  servers contain a top of rack switch, and multiple NIC-host
  simulators connecting to the top of rack switch. The Top of rack
  switch connects to the core switch through the proxy.}%
\end{figure*}

\subsubsection{Scaling Out}\hfill\\
We now move on to \sysname simulations running across
multiple physical hosts, using our RDMA and TCP proxies
(\autoref{fig:largescale-config}).

\paragraph{Overhead of distributed simulation.}
First we compare performance for local simulations to equivalent
distributed simulations with the \sysname proxies, to measure
overheads.
We use two qemu-kvm hosts running \texttt{netperf} connected to
\texttt{i40e} NICs which connect to the same switch.
Locally, this unsynchronized simulation yields a throughput of
4.4\,Gbps, and a latency of 71\,$\mu$s.
Next we distribute the simulation by running one pair of QEMU and NIC
on a second server and proxying the Ethernet connection to the switch
running locally.
With the sockets proxy the latency increases to 305\,$\mu$s and
throughput remains constant, and with RDMA both remain constant.
Next we measure simulation time for the same configuration but with
QEMU timing and gem5, and find that simulation time does not change
with either proxy.
We conclude that \emph{\sysname proxies are no bottleneck for
synchronized simulations.}

\paragraph{Large-scale memcache cluster.}
To evaluate scalability to larger systems, we next run multiple
distributed simulations ranging from 40 to 1000 simulated hosts, on 1
to 26 physical servers.
We run these simulations on Amazon ec2 \texttt{c5.metal} (spot)
instances, with 96 hyperthreads each, and 20\,Gbps network
connectivity in a single proximity placement group.
We simulate a varying number of racks of 40 hosts with \texttt{i40e}
NICs and a top of rack (ToR) switch each, that then connect to a
single core switch, as shown in \autoref{fig:largescale-config}.
We assign the core switch and each rack to a dedicated server.
A separate sockets proxy pair (Amazon ec2 does not offer RDMA)
connects each ToR to the core switch.
We run \texttt{memcached} on half of the hosts in each rack, and the
\texttt{memaslap} client on the other half.
Each client randomly connects to the 20 servers on the same rack, and
to 20 random servers in other racks.

\autoref{fig:large-scale} shows the measured simulation time for 10\,s
of virtual time as we increase the number of hosts.
From one rack and 40 hosts to 25 racks and 1000 hosts,
simulation time with gem5 hosts increases by 13.8\% from 15.5\,h,
to 17.6\,h.
With QEMU-timing, simulation time increases from 2.2\,h to 5.6\,h by
2.5$\times$.
With profiling we found the cause to be QEMU's dynamic binary
translation.
When an instance misses in its code cache and has to recompile a
block, the instance blocks for a while.
While rare, at scale these occurrences grow more frequent, and slow down
other hosts due to synchronization.
We conclude that \emph{\sysname scales to simulate systems with 100s
of hosts.}

\subsection{\sysname is Accurate}
\label{ssec:eval:accurate}
We now show \emph{\sysname Ethernet and PCIe interfaces accurately
connect and synchronize simulators}.
For Ethernet, we first run a pure ns-3 simulation of two communicating
nodes connected by a network link with our default parameters, and
log packet timestamps on each node.
Next, we repeat the experiment with two ns-3 instances each containing
one node and a \sysname Ethernet adapter, and connect the two.
For PCIe, we run two gem5 instances running netperf with the built-in
\texttt{e1000} NIC connected through the \sysname Ethernet adapter to a
switch.
We rerun this experiment with our standalone version of
gem5's \texttt{e1000} connected to both simulators through the \sysname
PCIe adapter.
In both cases we find that the timestamped logs match \textit{exactly},
demonstrating the correctness of our synchronization.

\subsection{\sysname is Deterministic}
\label{ssec:eval:deterministic}
Finally, we verify that \emph{\sysname simulations with deterministic component
simulators yields deterministic end-to-end simulations}.
To this end, we have repeated the two configurations combining only
deterministic simulators in \autoref{tab:modcombo} 5 times on
different machines.
We then compared event timestamps in the simulation logs and found
that they match \textit{exactly}.
\section{\sysname for System Evaluation}
Finally, we show \sysname end-to-end simulations can aid
evaluation, by providing more visibility and control than a
physical testbed, and by accurately simulating unavailable hardware.

\subsection{Use-Case: NIC Hardware Architecture}
\label{ssec:eval:corundum}
Using Corundum as an example, we show that \sysname simulations can
provide insights that are challenging to obtain from physical
testbeds.
The original Corundum evaluation shows significantly lower throughput
for a 1500B MTU compared to the ConnectX-5 NIC they compare to.
While developing our Corundum simulators, we found the root cause reason for
this.
Corundum relies on reading the head index registers of receive descriptor queues
to identify new entries, while for most other NICs, drivers instead directly
poll descriptors in memory.
MMIO reads stall the processor until the device returns a result,
while with DDIO descriptor reads typically hit in the L3 cache.
For CPU-bound workloads this degrades performance.

\paragraph{Leveraging simulation visibility \& flexibility.}
Our debugging effort was greatly facilitated by the simulator logs provided by
\sysname.
Synchronized simulations can produce detailed logs without affecting
system behavior.
We leveraged this to trace PCI activity, NIC activity, and CPU
activity, and combined those into an end-to-end view of the RPC latency.
We further confirm this by doubling the simulated PCIe latency to
1\,$\mu$s in gem5 with the Corundum and Intel behavioral simulators.
When PCIe latency doubles, Corundum throughput reduces by 21.2\%,
while the Intel NIC throughput remains unchanged.
Our experience demonstrates that \emph{simulators can offer greater visibility
and the flexibility to change key parameters that are fixed in physical
systems.}

\subsection{Use-Case: In-Network Processing}
\label{ssec:eval:nopaxos}
Work leveraging programmable switches for application acceleration
requires end-to-end measurements for a meaningful evaluation.
However, many of these works rely on functionality not (yet) available
in off-the-shelf hardware at publication time.
We use Network-Ordered Paxos (NOPaxos)~\cite{li:nopaxos} as an example
to demonstrate that \sysname can serve as a virtual testbed for such
systems.
NOPaxos introduces a new network-level primitive, the Ordered
Unreliable Multicast (OUM), which requires a single sequencer device
in the network.
Implementing the sequencer in a programmable network switch offers the best
performance.
However, as the required network hardware was not yet available, the authors
relied on sequencer emulation on a network processor or an end-host
implementation.
We implement switch support for OUM both in ns-3 and the now available Tofino
simulator, and combine them with gem5 and the Intel NIC.
On the simulated hosts, we run the unmodified NOPaxos open source code.

\paragraph{Reproducing results.}
We use \sysname to simulate two NOPaxos configurations: a P4 switch sequencer
running on Tofino, and an end-host sequencer implementation.
Similar to the original work, we also simulate the classic Multi-Paxos state
machine replication protocol.
We compare the throughput-latency curves (\autoref{fig:nopaxos-seq}) to figure 6
in the NOPaxos paper, where the switch sequencer configuration of NOPaxos achieves a
latency of 110\,$\mu$s, while the end-host sequencer configuration has ~35\%
higher latency; both configurations achieve similar throughput (230\,K/s).
The original paper also shows that NOPaxos (switch sequencer) achieves a
370\% increase in throughput and a 54\% reduction in latency compared to
Multi-Paxos.
In \sysname we find a lower baseline latency of 43\,$\mu$s for the switch
sequencer setup, and ~23\% higher latency for the end-host sequencer
configuration.
This is expected as the authors used a slower network processor
to emulate switch functionality.
We also find that both systems saturate at the same throughput of 78\,K/s.
The lower throughput is because we are measuring on a single-core
host, where application and packet processing share a core.
We confirmed this in a physical testbed by disabling all but one core,
and measured throughput within 10\%.
When comparing to Multi-Paxos running in \sysname, NOPaxos with switch sequencer
attains a 270\% throughput increase and 40\% latency reduction.
We conclude that \emph{\sysname can accurately evaluate in-network processing systems.}

\section{Looking Forward}
\label{sec:discussion}

\paragraph{Validation.}
To obtain representative simulation results, users have to
validate simulators and configuration parameters against physical
testbeds.
While \sysname cannot avoid this, we argue that our approach can
reduce validation effort.
Instead of validating each combination of simulators, components can
be validated individually and then composed.
This enables users to combine previously validated component
configurations into a full system.
We propose a public repository of validated component simulator
configurations to simplify re-use.
To ensure validity of these configurations over time, we imagine a
continuous-integration system, periodically re-running configurations and
recording the results.

\paragraph{Beyond networking.}
While we evaluate \sysname for network systems, our approach
generalizes beyond networking.
We have already demonstrated that our PCIe interface can support an
NVMe simulator.
Going forward, simulating PCIe attached accelerators, which are also
attracting growing interest in our community, should not require
changes to \sysname.
\sysname can also be easily extended with additional components or
interfaces, such as CXL~\cite{spec:cxl}.
We expect the emergence of further use-cases as architecture and
systems researchers continue to investigate specialized hardware.

\paragraph{Evaluating ASIC designs.}
Finally, we see evaluation of systems that include new ASIC components
as a driving use-case in the future.
While small ASIC designs with lower clock rates can often be evaluated
in physical testbeds with FPGAs, this is not possible for larger
designs or designs with fast clock speeds.
\sysname, on the other hand, can simulate ASIC RTL with arbitrary
frequencies, although FPGA accelerated RTL
simulations~\cite{karandikar:firesim} may be required for manageable
simulation times.
\section{Related Work}

\paragraph{Parallel \& distributed simulation.}
dist-gem5~\cite{mohammad:distgem5} and pd-gem5~\cite{alian:pd-gem5}
connect multiple gem5 instances for parallel and distributed
simulations and synchronize with global barriers.
Graphite~\cite{miller:graphite} also parallelizes a multi-core
simulation across cores and machines, but uses approximate
synchronization where causality errors are possible.
Similar to gem5, Simics~\cite{magnusson:simics} also supports full
system simulation and runs unmodified operating systems and
applications, and multiple Simics processes can be connected to
simulate networked systems.
\sysname connects multiple different simulators together using
fixed interfaces, and synchronizes them accurately with a
synchronization protocol that leverages the simulation structure.

ns-3 adds support for distributed simulation in version 3.8~\cite{dist-ns3}.
It uses a similar conservative look-ahead protocol with explicit
synchronization for correctness,and relies on the Message Passing
Interface (MPI) to connect multiple ns-3 processes.
MPI decouples ns-3 from the choice of message transport, directly
supporting distributed simulations over various interconnects, but
incurs the cost of this abstraction in every process.
\sysname instead closely couples synchronization and adapters to
our optimized shared memory queues (implementation is inlined from
shared headers), minimizing communication overhead in
simulator adapters.
\sysname scales out through proxies that decouple individual
simulators from the choice and overhead of distributed
transport (RDMA, sockets), at the cost of typically one core per
physical simulation host.

\paragraph{Co-simulation of multiple simulators.}
gem5 supports the integration of systemC
code~\cite{menard:gem5systemc} to implement hardware models, by
linking them into the gem5 binary and embedding the systemC event loop
with the gem5 event loop.
\sysname instead interconnects multiple heterogeneous simulators with
potentially completely different simulation models.
The Structural Simulation Toolkit (SST)~\cite{rodrigues:sst} is a
modular simulation framework for HPC clusters, uses a parallel
discrete event simulation with global epoch synchronization, and
defines common interfaces to link in various \textit{component}
simulators.
Unlike \sysname, SST requires deep integration of simulators into one
simulation loop resulting in integration challenges.
SST does also not define fixed component interfaces for specific
components, instead compatibility is up to individual simulators.

\paragraph{Full system emulation.}
Prior work on emulation has provided a path closer to end-to-end
evaluation without matching physical testbeds.
Mininet~\cite{lantz:mininet} emulates network topologies and hosts
through Linux networking and container features, running real
applications and using the host kernel for protocol processing.
ns-3 direct code execution (DCE)~\cite{tazaki:dce} integrates a
Linux Kernel instance as a libOS into ns-3 and connects its
network interface to ns-3 topologies.
Both systems offer lower run-times compared to \sysname, but at the cost of not
modeling low-level details, such as caches or PCIe interactions with devices, and
other bottlenecks on the physical system.
Finally, other work has relied on emulating NIC or switch
functionality on dedicated processors, while running the rest of the
system natively~\cite{kaufmann:flexnic,li:nopaxos}.
Simulations incur higher run-times but can control the level of details in the
model, and enable adjustment of relative performance of components by operating
on virtual time.

\section{Conclusion}

We described and evaluated \sysname, a novel modular framework
enabling full end-to-end simulation of network systems by combining
multiple tried-and-true simulators for different system components.
\sysname is fast and scalable, and accurately and deterministically
connects and synchronizes simulators.
We also demonstrated \sysname can replicate key findings from prior
work, including congestion control, in-network compute, and NIC
hardware architecture.
End-to-end simulations are a valuable tool for systems research,
especially in the era of specialized hardware.
 \makeatletter
\if@ACM@anonymous
\else
\section*{Acknowledgments}
We would like to thank the anonymous reviewers for their comments and feedback,
and the anonymous artifact evaluation committee for reviewing our artifact.
We also thank Jeff Mogul, Peter Druschel, Simon Peter, Trevor E. Carlson, Aastha
Mehta, Ming Liu, Katie Lim, Pratyush Patel, for their input on earlier drafts of this paper.
Keon Jang contributed the dctcp experiment idea and physical testbed
implementation, and joined many discussions on \sysname.
We thank Jonas Kaufmann for his help with preparing our artifact and open source
release, and Zhiqiang Xie for profiling \sysname.
Finally, we thank Huaicheng Li for help with integrating FEMU, and Tao
Wang and Anirudh Sivaraman for help with Menshen.
Jialin Li is supported by a MOE Tier 1 grant A-0008452-00-00 and a ODPRT grant
A-0008089-00-00.
 \fi
\makeatother

\label{page:lastbody}

\bibliography{paper,papers,strings,defs}

\clearpage
\appendix
\noindent
\emph{Appendices are supporting material that has not been
peer-reviewed.}

\section{Appendix}

\subsection{Modular Simulation Orchestration}
\label{ssec:appendix:orchestration}
Finally, an operational challenge arises for running simulations with \sysname.
Because we design \sysname without any centralized control, a simulation
consists entirely of interconnected component simulators.
Thus, to run a complete end-to-end simulation, a user has to start each
individual component simulator, while providing unique paths for the Unix
sockets and shared memory regions for each channel.
While this is manageable with very small simulations, the complexity rapidly
grows with simulation size, along with the additional challenges of cleanup,
collecting simulation logs, and monitoring for crashes.
An additional challenge, especially when running multiple simulations in
parallel, is that performance drastically degrades when overcommitting cores or
memory.
\sysname addresses both challenges with an orchestration framework for
assembling, running, and, if necessary, scheduling simulations.

\begin{figure}[h]%
\begin{lstlisting}[language=Python]
from simbricks import *
for rate in [10, 100, 200, 500, 1000]:
  e = Experiment('udp-' + str(rate))
  net = SwitchBM(e)

  s = Gem5Host(e, 'server')
  s.nic = I40eNIC(e)
  s.node_config = I40eLinuxNode()
  s.node_config.ip = '10.0.0.1'
  s.node_config.app = IperfUDPServer()

  c = Gem5Host(e, 'client')
  c.nic = I40eNIC(e)
  c.node_config = I40eLinuxNode()
  c.node_config.ip = '10.0.0.2'
  c.node_config.app = IperfUDPClient()
  c.node_config.app.server = '10.0.0.1'
  c.node_config.app.rate = rate

  experiments.append(e)
\end{lstlisting}\vspace{0.5em}%
\caption{An example of a simulation configuration in the \sysname
  orchestration framework.}%
\label{fig:orchestration}%
\Description{This figure shows pseudo-code but the text is included as
  in the PDF and as such should be accessible.}%
\end{figure}

Similar to other simulators with modular configuration we also implement our
orchestration in a scripting language.
The \sysname orchestration framework is designed as a collection of python
modules, and simulation experiments can be assembled by relying on arbitrary
python features.
In addition to the previously mentioned tasks, we also integrate functionality
to automatically generate customized disk images for host simulators, \eg with
different IP address configurations or to run applications with separate
parameters in individual hosts.
In \autoref{fig:orchestration} we show an example script.

\subsection{Inter-Simulator Message Transport}
\label{sec:appendix:shm}

\begin{figure}%
\begin{algorithmic}[0]%
  \State rxQueue, rxLen $\gets$ \Call{MapQueue}{rx}
  \State rxHead $\gets 0$
  \State txQueue, txLen $\gets$ \Call{MapQueue}{tx}
  \State txTail $\gets 0$
  \State
  \Procedure{PollMsg}{}
    \State msg $\gets$ \&rxQueue[rxHead]
    \While{msg->owner $\ne$ \texttt{CONSUMER}}
      \State \Call{Spin}{}
    \EndWhile
    \State \Call{ReadMemoryBarrier}{}
    \State rxHead $\gets$ (rxHead + 1) \% rxLen
    \State \Return{msg}
  \EndProcedure
  \Procedure{ReleaseMsg}{msg}
    \State msg->owner $\gets$ \texttt{PRODUCER}
  \EndProcedure
  \Procedure{AllocMsg}{}
    \State msg $\gets$ \&txQueue[txTail]
    \While{msg->owner $\ne$ \texttt{PRODUCER}}
      \State \Call{Spin}{}
    \EndWhile
    \State txTail $\gets$ (txTail + 1) \% txLen
    \State \Return{msg}
  \EndProcedure
  \Procedure{EnqueueMsg}{msg}
    \State \Call{WriteMemoryBarrier}{}
    \State msg->owner $\gets$ \texttt{CONSUMER}
  \EndProcedure
\end{algorithmic}%
\caption{\sysname multi-core shared memory message passing queue.
  \textsc{ReadMemoryBarrier} and \textsc{WriteMemoryBarrier} are compiler
  barriers to prevent re-ordering during optimization.}%
\label{fig:shm-queue}%
\Description{This figure shows pseudo-code but the text is included as
  in the PDF and as such should be accessible.}%
\end{figure}

\autoref{fig:shm-queue} shows pseudocode for the \sysname queue implementation.
To enable zero-copy implementation in simulators producer and consumer each have
separate functions for getting access to an available queue slot,
\textsc{PollMsg} for the consumer and \textsc{AllocMsg} for the producer, and
then releasing in when processing is complete, \textsc{ReleaseMsg} for the
consumer and \textsc{EnqueueMsg} for the producer.
The consumer uses its local head pointer to determine the slot the next message
is or will be in and then checks the type and ownership byte, re-trying if the
slot is marked by as owned by the producer.
After the consumer completes processing a message it marks the message as owned
by the consumer.
Symmetrically, the producer uses its local tail pointer to determine the slot
for the next message, if necessary waits until the slot is marked as
producer-owned, and resets the ownership bit to consumer after it places the
message in the slot.
Compiler memory barriers are necessary to prevent the compiler from reordering
memory accesses across accesses to the ownership bit, but with the strong X86
memory model no CPU memory barriers are necessary.

\subsubsection{Coherence Behavior}
To understand the performance properties, consider three key cases,
the queue is empty, the queue is full, and the queue is neither empty
nor full.
When the queue is empty, the consumer will spin on the last cache
line, which will be in the local L1 after the first access, and only
incurs an additional when the producer updates that cache line.
When the queue is full, the producer similarly waits for the next slot
to free up with the same coherence behavior.
Finally, when neither is the case, the consumer immediately finds a
message when polling and incurs a necessary miss that will fetch the
message.
Further, the CPU hardware prefetcher will likely already fetch the
next message as they are laid out sequentially in memory, thereby
avoiding a demand miss (but of course incurring the same coherence
traffic).
The producer does have to read the ownership flag incurring a miss,
but also immediately finds the empty slot, and the same prefetcher
behavior applies.

\subsection{\sysname Implementation Effort}
\label{ssec:apppendix:implcode}
\begin{table}%
  \centering%
  \begin{tabular}{ c c c }
      \toprule
      & \textbf{\sysname Component} & \textbf{Lines} \\
      \midrule
      \multirow{4}{50pt}{\sysname\newline core}
      & Message transport library & 1411 \\
      & NIC behavioral model library & 715 \\
      & Distributed simulation proxy & 2080 \\
      & Runtime orchestration & 2102 \\
      \midrule
      \multirow{2}{50pt}{Host simulators}
      & gem5 integration & 1265 \\
      & QEMU integration & 676 \\
      \midrule
      \multirow{4}{50pt}{NIC simulators}
      & Corundum Verilator & 1315 \\
      & Intel i40e model & 2900 \\
      & Corundum model & 911 \\
      & gem5 e1000 model & 2952 \\
      \midrule
      \multirow{5}{50pt}{Network simulators}
      & ns-3 integration & 158 \\
      & OMNeT++ integration & 208 \\
      & Tofino simulator integration & 330 \\
      & Ethernet switch model & 399 \\
      & Menshen RMT Verilator & 391 \\
      & Packet generator & 415 \\
      \midrule
      \multirow{1}{50pt}{Dev sims.}
      & FEMU SSD integration & 1005 \\
      \bottomrule\\
  \end{tabular}%
  \caption{Lines of code for the various components in \sysname, excluding
  blank lines and comments. For integrated simulators we only count
  adapter code.}%
  \label{tab:impl}%
\end{table}

\autoref{tab:impl} shows a per-component breakdown of the implementation effort
for \sysname, listing the number of lines of code.

\subsection{Performance for \sysname Configurations}
\label{ssec:appendix:simcombos}
\begin{table}%
\centering%
\begin{tabular}{lllrrrc}%
  \toprule
    \multicolumn{3}{c}{Simulators} & & &
    \multicolumn{1}{c}{Sim.}
    \\
    Host  & NIC & Net & T'put & Latency &
    \multicolumn{1}{c}{Time} &
    Det. \\
    \midrule
  QK & IB & SW & 4.37\,G & 71\,$\mu$s & 00:00:32 \\
  QK & IB & NS & 409\,M & 141\,$\mu$s & 00:00:32 \\
  QK & IB & TO & 1.92\,M & 6.6\,ms & 00:00:33 \\
  QK & CB & SW & 1.84\,G & 211\,$\mu$s & 00:00:29 \\
  QK & CB & NS & 429\,M & 294\,$\mu$s & 00:00:30 \\
  QK & CB & TO & 2.18\,M & 6.7\,ms & 00:00:33 \\
  QK & CV & SW & 81\,M & 3.4\,ms & 00:00:31 \\
  QK & CV & NS & 82\,M & 3.4\,ms & 00:00:32 \\
  QK & CV & TO & 2.31\,M & 23\,ms & 00:00:33 \\
\midrule
  QT & IB & SW & 8.85\,G & 17\,$\mu$s & 01:05:03 & (\checkmark) \\
  QT & IB & NS & 8.88\,G & 17\,$\mu$s & 01:06:43 & (\checkmark) \\
  QT & CB & SW & 3.74\,G & 28\,$\mu$s & 01:00:24 & (\checkmark) \\
  QT & CB & NS & 3.74\,G & 28\,$\mu$s & 00:59:41 & (\checkmark) \\
  QT & CV & SW & 6.55\,G & 32\,$\mu$s & 04:13:10 & (\checkmark) \\
  QT & CV & NS & 6.39\,G & 32\,$\mu$s & 04:13:13 & (\checkmark) \\
  G5 & IB & SW & 8.84\,G & 20\,$\mu$s & 12:51:41 & \checkmark \\
  G5 & IB & NS & 8.92\,G & 20\,$\mu$s & 12:49:46 & \checkmark \\
  G5 & CB & SW & 3.05\,G & 33\,$\mu$s & 09:20:48 & \checkmark \\
  G5 & CB & NS & 3.06\,G & 33\,$\mu$s & 09:26:13 & \checkmark \\
  G5 & CV & SW & 6.70\,G & 37\,$\mu$s & 10:23:26 & \checkmark \\
  G5 & CV & NS & 6.43\,G & 37\,$\mu$s & 10:21:28 & \checkmark %
 \\
    \bottomrule\\
\end{tabular}%
\caption{Performance for combinations of some of our component
    simulators.
    Checkmarks mark deterministic combinations.
    Host: QK is QEMU with KVM (functional simulation), QT
    is QEMU with timing, and G5 is gem5.
    NIC: IB is the Intel behavioral model, CB the Corundum
    behavioral model, and CV the Corundum verilator model.
    Network: SW is the switch behavioral model, NS is ns-3, TO is the
    Tofino model.
    }%
\label{table:crossproduct}%
\end{table}

\autoref{table:crossproduct} contains a cross-product of different simulators in
\sysname for host, NIC, and the network. This is an extended version
of \autoref{tab:modcombo} with the same experimental setup.
Note that with recent versions of QEMU we have found QEMU + timing
(QT) no longer to be fully deterministic and have instead observed
minor variations in simulation results.

\clearpage

\section{Artifact Appendix}

\subsection*{Abstract}

The \sysname artifact comprises two components, the source code of the
main simulator, and paper-specific parts (artifact scripts,
documentation, and data) to replicate the results in this
paper.

\subsection*{Scope}
Users interested in using \sysname in their work should refer to the
former, as this will continue to evolve over time, while the latter
remains stable (modulo bug fixes) to ensure reproducible results.

The artifact scripts can run all major and minor experiments in
the paper, except for the physical testbed baseline for the dctcp
experiments.
For deterministic simulations, results should be exactly reproducible.
Other measurements, especially simulation times, will vary based on
the hardware, but should be approximately reproducible on
similar hardware to what we describe.

\subsection*{Contents}
The artifact contains everything required to reproduce the results in
the paper: source code, instructions for building and running
\sysname, scripts for running experiments, and plotting scripts for
the graphs in the paper. We also include most of the execution logs we
generated for the experiments in this paper.

\subsection*{Hosting}
Both the main \sysname repo and the artifact package are hosted on
GitHub:
\begin{itemize}
  \item \textbf{Main \sysname source:} \newline
    \url{https://github.com/simbricks/simbricks}
  \item \textbf{Artifact package:} \newline
    \url{https://github.com/simbricks/sigcomm22-artifact}
\end{itemize}

For both we have tagged the version submitted for evaluation with
\texttt{sigcomm22-ae-submission}, and a stable version potentially
receiving bug-fixes will remain in the \texttt{sigcomm22-ae} branch.
The \texttt{main} branch will evolve and might contain breaking
changes.

We have also built docker specifically for the artifact that we link to in the
artifact README file.

\subsection*{Requirements}
The precise hardware requirements for each experiment vary
significantly and are detailed in the artifact repository. All
non-distributed experiments only require a single machine, but require
sufficient processor cores (varies per experiment up to 44). The
largest experiments also require around 192\,GB of RAM.

We have tested \sysname on Linux. The specific software dependencies
are provided by the documentation in the artifact repo.

\label{page:last}
\end{document}